\documentclass[11pt,a4paper]{article}
\usepackage{jstyle}
\usepackage[utf8]{inputenc}
\usepackage{framed}
\usepackage{youngtab}
\usepackage{amsmath}
\usepackage{enumitem}
\usepackage{color}
\usepackage{graphicx}
\usepackage{hyperref}

\makeatletter
\def\@xfootnote[#1]{%
  \protected@xdef\@thefnmark{#1}%
  \@footnotemark\@footnotetext}
\makeatother

\newcommand*\longhookrightarrow{\ensuremath{\lhook\joinrel\relbar\joinrel\rightarrow}}

\newcommand{\bi}{\begin{itemize}}
\newcommand{\ei}{\end{itemize}}
\newcommand{\bea}{\begin{align}}
\newcommand{\eea}{\end{align}}
\newcommand{\be}{\begin{equation}}
\newcommand{\ee}{\end{equation}}

\makeatletter
\renewcommand*\env@matrix[1][\arraystretch]{%
  \edef\arraystretch{#1}%
  \hskip -\arraycolsep
  \let\@ifnextchar\new@ifnextchar
  \array{*\c@MaxMatrixCols c}}
\makeatother

\author{\large{Charlotte Sleight}}

\affiliation{\large{Universit\'e Libre de Bruxelles
and International Solvay Institutes \vspace*{0.1cm}\\ 
ULB-Campus Plaine CP231, 1050 Brussels, Belgium}\\}

\affiliation{\large{Max-Planck-Institut f\"ur Physik\\
F\"ohringer Ring 6, 80805 Munich, Germany}}

\emailAdd{charlotte.sleight@gmail.com}

\title{\huge{Higher Spin Holography}\\ \vspace*{0.2cm}
\Large{Lecture Notes for the XII Modave School in Mathematical Physics}\footnote[$\dagger$]{11-17th September 2016, Modave, Belgium. Web page: \url{http://www.ulb.ac.be/sciences/ptm/pmif/Rencontres/ModaveXII}.}}

\abstract{{\large These notes were prepared for the introductory lectures on Higher Spin Holography presented at the
Twelfth Modave Summer School in Mathematical Physics 2016, aimed at Ph.D. students and researchers new to this topic. Contents:\\

\noindent
1. Higher spin particles in AdS\\
2. The AdS/CFT correspondence and higher spins\\
3. The ambient space formalism\\
4. Witten diagrams for higher spin fields / higher spin interactions from CFT}}

\begin{document}

\maketitle

\section*{Acknowledgements}

I thank the organisers of the Twelfth Modave Summer School in Mathematical Physics for the kind invitation to present this material and, together with the other participants and lecturers, for the stimulating discussions + questions.
The original results presented in the final section of these notes were obtained as part of collaborations with Xavier Bekaert, Johanna Erdmenger, Mitya Ponomarev \cite{Bekaert:2014cea,Bekaert:2015tva} and Massimo Taronna \cite{Sleight:2016dba}, whom I thank for enlightening discussions over the years. I am also indebted to Massimo Taronna for constructive comments on the draft.

\newpage

\section*{Introduction}

The study of higher-spin gauge theories has a long history. Since the early works in the 1930's of Majorana \cite{Majorana:1968zz}, Dirac \cite{10.2307/96758}, Fierz \cite{Fierz:1939zz} together with Pauli \cite{Fierz:1939ix}, and Wigner \cite{Wigner:1939cj}, by now the free propagation of higher-spin gauge fields is rather well understood (for instance see: \cite{Fronsdal:1978rb,Fang:1978wz,LABASTIDA1986101,PhysRevLett.58.531,Labastida:1987kw,Metsaev:1995re,Segal:2002gd,Alkalaev:2003qv,Alkalaev:2003hc,Germani:2007em,Ferrara:2011km,Kulaxizi:2014yxa,Metsaev:2014iwa,Nutma:2014pua,Rahman:2016tqc}). On the other hand, the question of constructing consistent interactions among them is a highly non-trivial one (for a review see \cite{Bekaert:2010hw}). 

One of the main motivations for studying higher-spin gauge theories is the on-going quest for a UV-complete theory of gravity. Indeed, upon the addition of higher-derivative counter-terms to the Einstein-Hilbert action,\footnote{For example, at one-loop one includes the Gauss-Bonnet term.} to avoid violations of causality at the classical level we are led to introduce an infinite tower of massive particles of spins $s>2$ \cite{Camanho:2014apa,DAppollonio:2015fly}. One may then expect an underlying \emph{higher-spin} symmetry principle governing the high-energy behaviour of the theory, whose spontaneous breaking would generate the lower energy spectrum of massive higher-spin states. This picture was also motivated from a String Theory perspective by Gross \cite{Gross:1988ue} in the 1980's.

Higher-spin gauge theories have generated an increased interest in the last two decades, owing in particular to their role in the celebrated AdS/CFT correspondence \cite{Maldacena:1997re,Gubser:1998bc,Witten:1998qj}. Theories of higher-spin gauge fields on anti-de Sitter backgrounds have been conjectured to be dual to very simple, free, Conformal Field Theories \cite{HaggiMani:2000ru,witten60thschw,Sezgin:2002rt,Klebanov:2002ja}. This has the potential to provide a powerful framework to acquire a deeper understanding of AdS/CFT, and also into higher-spin gauge theories themselves.

In these notes we will focus on the latter. 
In particular, we review some recent efforts \cite{Bekaert:2014cea,Bekaert:2015tva,Sleight:2016dba,Sleight:2016hyl} aimed at using holography to study interactions of higher-spin particles.\footnote{For related works by other authors see \cite{Petkou:2003zz,Skvortsov:2015lja,Skvortsov:2015pea}.} To this end, we introduce some useful tools for computing tree-level amplitudes in AdS involving fields of arbitrary integer spin. These are underpinned by the so-called ambient space formalism introduced by Dirac in the 1930's \cite{Dirac:1936fq}, in which anti-de Sitter space is viewed as a one-sheeted hyperboloid embedded in a higher-dimensional flat space. In order to be self-contained, we also review relevant aspects of the AdS/CFT correspondence and the basics of higher-spin particles on AdS.

\section{Higher Spin Particles in AdS}

\subsection{The AdS Geometry and Isometry Group}
\label{subsec::adsgeom}
From the holographic view point taken in these lectures, we are interested in particles propagating on a $\left(d+1\right)$-dimensional anti-de Sitter (AdS$_{d+1}$) background, which can be regarded as the hyperboloid\footnote{The length scale $R$ is known as the AdS radius, which is related to the cosmological constant $\Lambda$ via 
\begin{equation}
    \Lambda = -\frac{d\left(d-1\right)}{2R^2} < 0.
\end{equation}}
\begin{equation}
    -X^2_0-X^2_{d+1}+\sum^{d}_{i=1} X^2_i = -R^2, \label{hyp}
\end{equation}
embedded in an \emph{ambient} $\left(d+2\right)$-dimensional flat space time with metric
\begin{equation}
    ds^2 = \eta_{AB}dX^AdX^B =  -dX^2_0-dX^2_{d+1}+\sum^{d}_{i=1} dX^2_i,
\end{equation}
where $\eta_{AB} = \text{diag}\left(-++\cdot\cdot\cdot+-\right)$ and $A, B = 0, 1, ...\,, d+1$. More concretely, denoting the intrinsic co-ordinates on our hyperboloid \eqref{hyp} by $x^\mu$ (with $\mu = 0,...,d$), we are making a smooth isometric embedding
\begin{equation}
i\; : \quad H_{d+1} \: \longhookrightarrow \: \mathbb{R}^{d+2} : \quad x^{\mu} \longmapsto X^{A}\left(x^{\mu}\right).
\end{equation}
From the above we can see that AdS$_{d+1}$ space is homogeneous and isotropic, with isometry group $SO\left(d,2\right)$. The corresponding algebra consists of the  $\frac{1}{2}\left(d+1\right)\left(d+2\right)$ generators
\begin{equation}\label{jab}
  i J_{AB} = -i J_{BA} = \left(X_A \frac{\partial}{\partial X^B}-X_B\frac{\partial}{\partial X^A}\right), 
\end{equation}
which satisfy the commutation relations
\begin{equation}
    \left[J_{AB}, J_{CD}\right] = i\left(\eta_{BC}J_{AD}+\eta_{AD}J_{BC}-\eta_{AC}J_{BD}-\eta_{BD}J_{AC}\right),\label{sod2}
\end{equation}
with $\eta_{AB} = \text{diag}\left(-++\cdot\cdot\cdot+-\right)$, known as `conformal signature'. For calculations it is often convenient to work in Euclidean AdS, which can be reached by instead working in an ambient space Lorentzian signature $\eta_{AB}=\text{diag}\left(-+\cdot\cdot\cdot+\right)$.

It is often convenient to use the following basis for the $\mathfrak{so}\left(d,2\right)$ generators \eqref{sod2},  
\begin{equation}
    M_{ij}=iJ_{ij}, \qquad P^\pm_i = J_{0i} \pm i J_{i(d+1)},  \qquad E = J_{0(d+1)},\label{confcomb}
\end{equation}
with commutators (all others are vanishing)
 \begin{eqnarray}\label{confalg1}
 &\left[E, P^{\pm}_i\right] = \pm P^\pm_i, \quad
     \left[M_{ij}, P^\pm_{k}\right] = \delta_{kj }P^\pm_{i} - \delta_{ki}P^\pm_j,\\ \label{confalg2}
    &  \left[M_{i j}, M_{k l}\right] = \delta_{i k}M_{j l} + \delta_{j l}M_{i k}-\delta_{j k}M_{i l} -\delta_{i l}M_{j k}, \quad
      \left[P^+_i,\,P^-_j\right] = 2M_{ij}-2\delta_{ij}E,
 \end{eqnarray}
  where $i,j=1,...,d$.

\subsection*{Euclidean AdS in Poincar\'e Co-ordinates}

For concreteness, the co-ordinate system we'll most often use is Euclidean AdS in Poincar\'e co-ordinates $x^\mu = \left(z,y^i\right)$. With this choice our hyperboloid \eqref{hyp} is parameterised by 
\begin{align}
    X^0\left(x\right) &= R\frac{z^2+y^2+1}{2z} \\
    X^{d+1}\left(x\right) &=R\frac{1-z^2-y^2}{2z} \\
    X^i\left(x\right) & = \frac{Ry^i}{z}. 
\end{align}
Pulling back the ambient metric $\eta_{AB}$ one recovers the AdS metric in Poincar\'e co-ordinates
\begin{equation}\label{poin}
    ds^2 = \left(\frac{\partial X^A}{\partial x^\mu}\frac{\partial X^B}{\partial x^\nu}\eta_{AB}\right) dx^\mu dx^\nu = \frac{R^2}{z^2}\left(dz^2+\delta_{ij}dy^idy^j\right),
\end{equation}
where we used the Lorentzian signature $\eta_{AB}=\text{diag}\left(-+\cdot\cdot\cdot+\right)$.

The usual AdS Killing tensors can be obtained in Poincar\'e co-ordinates by noting that
\begin{equation}
\eta^{AB} \frac{\partial}{\partial X^B} = g^{\mu \nu}  \frac{\partial X^A}{\partial x^\nu} \frac{\partial}{\partial x^\mu} - \frac{X^A}{R^2} X \cdot \partial_X,
\end{equation}
which gives
\begin{align}
 i J^{AB} & = g^{\mu \nu}\left(X^{A}\frac{\partial X^B}{\partial x^\nu}-X^{B}\frac{\partial X^A}{\partial x^\nu}\right)\frac{\partial}{\partial x^\mu} \\ \nonumber
 & = \frac{z^2}{R^2}\left[\left(X^{A}\frac{\partial X^B}{\partial z}-X^{B}\frac{\partial X^A}{\partial z}\right) \frac{\partial}{\partial z}+\delta^{ij}\left(X^{A}\frac{\partial X^B}{\partial y^j}-X^{B}\frac{\partial X^A}{\partial y^j}\right) \frac{\partial}{\partial y^i}\right].
\end{align}
For example, one recovers
\begin{equation}
 iE = J_{0(d+1)} = z \partial_z + y \cdot \partial_y.
\end{equation}

\subsection{The Conformal Boundary}
\label{subsec::confbo}

\begin{figure}[h]
  \centering
  \includegraphics[scale=0.4]{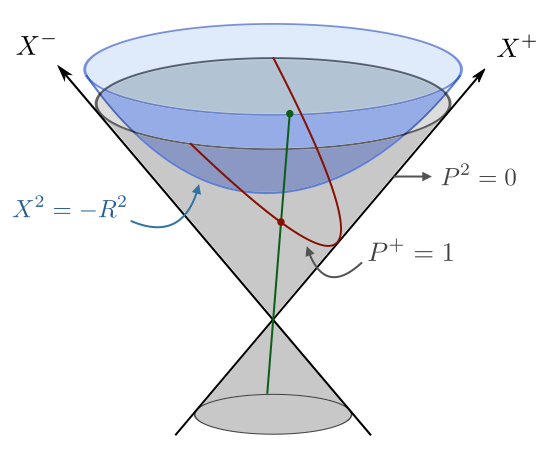}
 \caption{} \label{ambient}
\end{figure}

Towards the AdS$_{d+1}$ boundary, the hyperboloid \eqref{hyp} asymptotes to a light cone in the flat ambient space (see fig. 1). While this limit does not yield a well-defined boundary metric (see \eqref{poin} for $z \rightarrow 0$), we can obtain a finite limit by considering the projective cone of light rays with co-ordinates
\begin{equation}
  P^A \;\equiv \; \epsilon X^A, \qquad \epsilon \rightarrow 0.\label{limp}
\end{equation}
Since $X^2$ is fixed, these null projective co-ordinates satisfy
\begin{equation}
  P^2=0, \qquad P\;\sim\;\lambda P, \qquad \lambda \ne 0, \label{eclass}\end{equation}
where in general $\lambda$ depends on $P$. The right-most statement tells us that, being rays, $P$ and $\lambda P$ are identified. This quotienting of the null cone identifies the $P$ with $d$-dimensional Minkowski space (with a point at infinity added) or any conformally flat manifold.\footnote{Note that these redundant rescalings by $\lambda(y)$ are equivalent to sending $\epsilon \rightarrow \epsilon / \lambda$, so $P\rightarrow\lambda(y) P$ rescales the metric by an overall factor -- as per the definition of a conformal transformation. The redundancy can thus be used to determine the transformation properties of a function $f\left(P\right)$ under a dilatation. If $f$ has scaling dimension $\Delta$, then $f\left(\lambda P\right) = \lambda^{-\Delta}f\left(P\right)$.} 

In particular, the AdS boundary is parameterised by a Poincar\'e section of the null cone: $P^+ = P^{d+1}+P^0= \text{constant}$. With the gauge choice $P^+=1$, we have
\begin{equation}\label{sect}
  P^{0}\left(y\right)=\frac{1}{2}\left(1+y^2\right), \quad P^{d+1}\left(y\right)=\frac{1}{2}\left(1-y^2\right), \quad P^i\left(y\right) = y^i.
  \end{equation}
This is illustrated in fig. \ref{ambient}. The $SO\left(d,2\right)$ isometry of AdS acts on the co-ordinates $P$ as a group of conformal symmetries, with a given transformation relating the section \eqref{sect} to others with $dP^+ = 0$.

The usual conformal generators on the boundary are recovered from 
\begin{equation}\label{jabp}
  i J_{AB} = -i J_{BA} = \left(P_A \frac{\partial}{\partial P^B}-P_B\frac{\partial}{\partial P^A}\right), 
\end{equation}
and identifying the combinations \eqref{confcomb}. Like for the AdS Killing tensors in the previous section \S \ref{subsec::adsgeom}, one can use that
\begin{align}\nonumber
 \eta^{AB}\frac{\partial}{\partial P^B} & = \left(\frac{\partial P^A}{\partial y^i}\frac{\partial P^B}{\partial y^i}-Q^AP^B-Q^BP^A\right)\frac{\partial}{\partial P^B} \\ \nonumber
& = \delta^{ij}\frac{\partial P^A}{\partial y^i} \frac{\partial}{\partial y^j}-Q^A P \cdot \partial_P-P^A Q \cdot \partial_P,
\end{align}
where we employed \eqref{uplift} with $Q^A = \left(1,\mathbf{0},-1\right)$. This gives
\begin{align}
iJ^{AB} = \left(P^A \frac{\partial P^B}{\partial y^i} - P^B \frac{\partial P^A}{\partial y^i} \right)\frac{\partial}{\partial y^i}+\left(P^BQ^A-P^AQ^B\right) y \cdot \partial_y, 
\end{align}
where one notes that $P \cdot \partial_P = y \cdot \partial_y$. For example, for dilatations we recover
\begin{equation}
   i E = i J_{0\left(d+1\right)} = y \cdot \partial_{y}.\label{econf}
\end{equation}

\subsection{Unitary Irreducible Representations: Particles in AdS}
\label{subsec::uir}
All possible types of elementary particles are defined by  unitary irreducible representations (UIRs) of the space-time isometry. For studying the UIRs of the AdS isometry, it is useful to employ the basis \eqref{confalg2} for the $\mathfrak{so}\left(d,2\right)$ generators. Since we are interested in unitarity representations of ${\mathfrak so}\left(d,2\right)$, the required Hermiticity condition $J_{AB}^\dagger = J_{AB}$ translates into
  \begin{equation}
      E^\dagger = E, \qquad \left(P^\pm \right)^\dagger = P^\mp, \qquad M_{ij}^\dagger = -M_{ij}.   
  \end{equation}
  Generators $E$ and $M_{ij}$ comprise the maximally compact subgroup, $SO\left(2\right) \times SO\left(d\right)$, of $SO\left(d,2\right)$. $E$ gives rise to rotations in the purely time-like $\left(X_0,X_{d+1}\right)$ plane and therefore identified with the Hamiltonian for AdS physics. The $M_{ij}$ give rotations of $S^{d-1}$ and are angular momentum generators. The remaining non-compact generators $P^\pm$ \emph{raise} and \emph{lower} the energy eigenvalue by one unit respectively, as can be seen from the $\left[E, J^{\pm}\right]$ commutator in \eqref{confalg1}.
 
 UIRs of the AdS isometry group are thus labelled by the energy eigenvalue $\Delta$ and spin ${\underline s}$ of its ground state $|\Delta,{\underline s}\rangle$
 \begin{equation}
     E|\Delta,{\underline s}\rangle = \Delta |\Delta,{\underline s}\rangle, \qquad  P^- |\Delta,{\underline s}\rangle = 0,\label{gs}
 \end{equation}
 which forms a unitary module of the ${\mathfrak so}\left(2\right) \oplus {\mathfrak so}\left(d\right)$ maximally compact sub-algebra. The spin ${\underline s}$ characterises the ${\mathfrak so}\left(d\right)$ module, which is generically given by a collection of positive integers ${\underline s}=\left(s_1, ..., s_r\right)$ corresponding to the ${\mathfrak so}\left(d\right)$ Young diagram 
 \begin{equation}
 \begin{aligned}
&\begin{tabular}{|c|c|c|c|c|}\hline
   $\phantom{.}$&$\phantom{.}$&\multicolumn{2}{|c|}{$~~~~\cdots~~~~\cdots~~~~\cdots~~~~$}&\phantom{.}\\\hline
\end{tabular}~\,s_1\\[-4pt]
&\begin{tabular}{|c|c|c|c|c|}
   $\phantom{.}$&$\phantom{.}$&\multicolumn{2}{|c|}{$~~~~\cdots~~~~$}&$\phantom{.}$\\\hline
\end{tabular}~\,s_2\\[-4.3pt]
&\begin{tabular}{|c|}
   $~~~~\vdots~~~~\vdots~~~~\vdots~~~~$\\\hline
\end{tabular}\\[-4pt]
&\begin{tabular}{|c|c|c|}
   $\phantom{.}$&$\cdots$&$\phantom{.}$\\\hline
\end{tabular}~\,s_r\;,
 \end{aligned}
  \end{equation}
 with $s_1 \ge s_2 ... \ge s_r$. For the purpose of these lectures (and from this point onwards) we only consider totally symmetric spin-$s$ representations $V_s$, which concern only the single row Young diagrams ${\underline s} = \left(s,0,...,0\right)$,
 \begin{equation}
 \begin{aligned}
&\begin{tabular}{|c|c|c|c|c|}\hline
   $\phantom{.}$&$\phantom{.}$&\multicolumn{2}{|c|}{$~~~~\cdots~~~~\cdots~~~~$}&\phantom{.}\\\hline
\end{tabular}~\,s\,.
 \end{aligned}
  \end{equation}
Given the ground state \eqref{gs} of our spin-$s$ particle, we can construct excited states furnishing the representation Fock space by applying the raising operator $P^+$. The complete ${\mathfrak so}\left(d,2\right)$ module ${\cal D}\left(\Delta,s\right)$ is therefore spanned by states of the schematic form\footnote{In particular this is schematic for $l>0$, as, strictly speaking, the indices should be symmetrised in order to be irreducible $SO\left(d\right)$ tensors.}
\begin{equation}
    |\Delta,s\rangle_{n,l} = \left(P^+ \cdot P^+ \right)^n P^+_{i_1} ... P^+_{i_l}|\Delta,s\rangle, \quad n,\,l=0,1,2...\,,
\end{equation}
 with energy eigenvalue $\Delta+2n+l$. Note that  $|\Delta,s\rangle_{0,0}=|\Delta,s\rangle$.  

\begin{framed}
\noindent \underline{Exercise 1.1:} Quadratic Casimir\\

\noindent Casimir operators of a given Lie algebra are distinguished operators which commute with each generator. Their eigenvalues thus characterise the irreducible representations, taking the same value for any state in a given representation. 

The quadratic Casimir of the AdS$_{d+1}$ isometry algebra is given by 
\begin{equation}
    {\cal C}_2\left({\mathfrak so}\left(d,2\right)\right) \equiv \frac{1}{2}J_{AB}J^{AB} = E\left(E-d\right)+{\cal C}_2\left({\mathfrak so}\left(d\right)\right)-\delta^{ij}P^+_i P^-_j.
\end{equation}
Given that the ${\mathfrak so}\left(d\right)$ Casimir\footnotemark \,  ${\cal C}_2\left({\mathfrak so}\left(d\right)\right) =- \frac{1}{2}M_{ij}M^{ij}$ has eigenvalue
\begin{equation}
   \langle {\cal C}_2\left({\mathfrak so}\left(d\right)\right) \rangle = s\left(s+d-2\right),
\end{equation}
 on totally symmetric spin-$s$ representations $V_s$, show that
\begin{equation}
    \langle {\cal C}_2\left({\mathfrak so}\left(d,2\right)\right) \rangle = \Delta\left(\Delta-d\right) + s\left(s+d-2\right),
\end{equation}
for states in the module ${\cal D}\left(\Delta,s\right)$. 

Show also that
\begin{equation}
    {\cal C}_2\left({\mathfrak so}\left(d,2\right)\right) = R^2\Box_{\text{AdS}} + {\cal C}_2\left({\mathfrak so}\left(d,1\right)\right),
    \end{equation}
where $\Box_{\text{AdS}} = \nabla^{\mu}\nabla_{\mu}$ is the covariantised d'Alambertian operator in AdS and ${\cal C}_2\left({\mathfrak so}\left(d,1\right)\right)$ is the quadratic Lorentz Casimir in $(d+1)$-dimensions.
\end{framed}
 \footnotetext{Notice here the minus sign, since the generators $M_{ij}$ are anti-Hermitian, differing from the usual Hermitian generators $J_{ij}$ by a factor of $i$.}
 
 In the language of QFT, to our spin-$s$ particle on AdS is associated a rank $s$ field $\varphi_{\mu_1...\mu_s}$  which, as a carrier of ${\cal D}\left(\Delta,s\right)$, is totally symmetric and satisfies the Fierz-Pauli conditions
 \begin{subequations}\label{fp}
 \begin{eqnarray}
 &\left({\cal C}_2\left({\mathfrak so}\left(d,2\right)\right) - \langle {\cal C}_2\left({\mathfrak so}\left(d,2\right)\right) \rangle \right)\varphi_{\mu_1...\mu_s} = 0, \label{casi} \\ \label{divfp}
 & \nabla^{\mu_1}\varphi_{\mu_1...\mu_s}  = 0,\\
 & g^{\mu_1 \mu_2} \varphi_{\mu_1...\mu_s}  = 0.
 \end{eqnarray}
 \end{subequations}
 The final two conditions ensure that $\varphi_{\mu_1...\mu_s}$ sits in $V_s$, the totally symmetric irreducible spin-$s$ representation of $SO\left(d\right)$. From the first condition \eqref{casi} one deduces the equation of motion (see exercise 1.1 above)
 \begin{equation}\label{casieomfp}
     \left(\nabla^2-m^2_s\right)\varphi_{\mu_1...\mu_s} = 0, \quad \left(m_s R\right)^2 = \Delta\left(\Delta-d\right)-s.
 \end{equation}
 Recall that $R$ is the AdS radius.
 \subsection*{Unitarity Bounds and Higher Spin Gauge Fields}
 
 Let us emphasise that representations are only unitary for a certain range of $\Delta$. Outside of this, negative norm states appear in the Hilbert space. Unitarity bounds on $\Delta$ can be obtained by demanding positive norm for every state in the multiplet, which we detail below. 
 
It is sufficient to consider the norm of the first level descendants,
\begin{equation}
    \left( P^+_i | \Delta,s\rangle\right)^\dagger P^+_j | \Delta,s\rangle = \langle \Delta, s | P^-_i P^+_j | \Delta, s\rangle = 2\Delta \delta_{ij} - 2 \Sigma_{ij}, \label{1level} 
\end{equation}
 where we used the ${\mathfrak so}\left(d,2\right)$ commutator \eqref{confalg2} and $M_{ij}| \Delta, s\rangle^a = \left(\Sigma_{ij}\right){}^{a}{}_b | \Delta, s\rangle^b$, with $a,b$ indices for the $SO\left(d\right)$ representation $V_s$ of $| \Delta, s\rangle$. For unitarity we require that \eqref{1level} is positive definite, which implies
 \begin{equation}
     \Delta \geq \text{max. Eigenvalue}\left[\left(\Sigma_{ij}\right){}^{a}{}_b\right].
 \end{equation}
 The state $P^+_i | \Delta,s\rangle$ sits in the $V_1 \otimes V_s$ representation of $SO\left(d\right)$, where $V_1$ is the vector representation. The trick is to write
 \begin{equation}
     \left(\Sigma_{ij}\right){}^{a}{}_b = \frac{1}{2} (L^{kl})_{ij}\left(\Sigma_{kl}\right){}^{a}{}_b, \label{sig}
 \end{equation}
 where $(L^{kl})_{ij} = \delta^k_i\delta^l_j - \delta^k_j\delta^l_i$ is the spin generator in $V_1$. Then, regarding $L^{kl}$ and $\Sigma_{kl}$ as operators acting on $V_1 \otimes V_s$, \eqref{sig} becomes
 \begin{align}
     L^{kl}\Sigma_{kl} & = \frac{1}{2}\left(\left(L+\Sigma\right)^2-L^2-\Sigma^2\right)\\
     & =-\langle{\cal C}_2\left({\mathfrak so}\left(d\right)\right)\rangle\big|_{V_1 \otimes V_s}+\langle{\cal C}_2\left({\mathfrak so}\left(d\right)\right)\rangle\big|_{V_1}+\langle{\cal C}_2\left({\mathfrak so}\left(d\right)\right)\rangle\big|_{V_s}.
 \end{align}
 We see that the maximum Eigenvalue of $L \cdot \Sigma$ is dictated by the minimal quadratic ${\mathfrak so}\left(d\right)$ Casimir in $V_1 \otimes V_s$. By  decomposing\footnote{Note that this decomposition holds only for $s>0$. For $s=0$ we have $V_1 \otimes V_{s=0} = V_1$, leading to a modified bound, $\Delta \ge \frac{d}{2}-1$ or $\Delta = 0$, for $s=0$. This gives the Breitenlohner-Freedman bound \cite{Breitenlohner:1982bm}
 \begin{equation}
     m^2_0 > -\left(\frac{d}{2R}\right)^2,
 \end{equation}
 which tells us that fields are still stable in AdS even if they are a little bit tachyonic.}
  \begin{equation}
  \begin{alignedat}{2}
 &   \begin{tabular}{|c|}\hline
   $1$\\\hline
\end{tabular}\;\otimes\; 
\begin{tabular}{|c|c|c|c|c|}\hline
   $1$&\multicolumn{2}{|c|}{$~~~~\cdots~~~~$}&$s$\\\hline
\end{tabular}\\
& \hspace*{2cm} = \;\begin{tabular}{|c|c|c|c|c|}\hline
   $1$&\multicolumn{2}{|c|}{$~~~~\cdots~~~~$}&$s-1$\\\hline
\end{tabular}\;\oplus\;
\begin{tabular}{|c|c|c|c|c|}\hline
   $1$&\multicolumn{2}{|c|}{$~~~~\cdots~~~~$}&$s+1$\\\hline
\end{tabular}
\;\oplus\; \begin{tabular}{|c|c|c|c|c|}\hline
    $1$&\multicolumn{2}{|c|}{$~~~~\cdots~~~~$}&$s$\\\hline
\end{tabular}\:,\\[-4pt] \nonumber
&\hspace*{11.94cm}\begin{tabular}{|c|}\hline
   $\phantom{1}$\\\hline
\end{tabular}\,
 \end{alignedat}
 \end{equation}
 \noindent we see that this is given by $V_{s-1}$. We therefore obtain the bound
 \begin{align}
     \Delta \ge \frac{1}{2}\left(-\langle{\cal C}_2\left({\mathfrak so}\left(d\right)\right)\rangle\big|_{V_{s-1}}+\langle{\cal C}_2\left({\mathfrak so}\left(d\right)\right)\rangle\big|_{V_1}+\langle{\cal C}_2\left({\mathfrak so}\left(d\right)\right)\rangle\big|_{V_s}\right) = s+d-2.
 \end{align}
 While below this bound some states have negative norm, when it is saturated ($\Delta = s+d-2$) null states emerge, which are orthogonal to all states in the Hilbert space. Such states hence form an invariant sub-module which should be quotiented out, corresponding to the emergence of a gauge symmetry. Indeed, one may verify the Fierz system \eqref{fp} for $\Delta = s+d-2$:
 \begin{subequations}\label{fpgauge}
 \begin{eqnarray}
 & \left(R^2\nabla^2-(s+d-2)\left(s-2\right)+s\right)\varphi_{\mu_1...\mu_s} = 0, \label{casi2} \\ \label{divg2}
 & \nabla^{\mu_1}\varphi_{\mu_1...\mu_s}  = 0,\\ \label{trg2}
 & g^{\mu_1 \mu_2} \varphi_{\mu_1...\mu_s}  = 0,
 \end{eqnarray}
 \end{subequations} 
 is invariant under the gauge transformation 
 \begin{equation}\label{linhsg}
    \delta_{\xi} \varphi_{\mu_1...\mu_s} = \nabla_{\left(\right.\mu_1}\xi_{\mu_2...\mu_s\left.\right)},
 \end{equation}
 where the symmetric and traceless gauge parameter $\xi$ is on-shell:
\begin{align}\label{xifp}
  \left(R^2\nabla^2-\left(s-1\right)\left(s+d-3\right)\right)\xi_{\mu_1...\mu_{s-1}}&=0, \\ \nabla^{\mu_1}\xi_{\mu_1...\mu_{s-1}}&=0, \\ g^{\mu_1\mu_2}\xi_{\mu_1...\mu_{s-1}}&=0.
\end{align}
In contrast to our intuition from flat space, we see that gauge fields in AdS have a mass owing to the background curvature.

\begin{framed}
\noindent \underline{Exercise 1.2:} Generating Functions\\

\noindent For manipulations of higher-rank tensors, it is useful to employ an operator notation. Fields are represented by generating functions, which for totally symmetric spin-$s$ representations read
\begin{equation}
\varphi_{\mu_1...\mu_s}\left(x\right) \: \longrightarrow \: \varphi_s\left(x,u\right) = \frac{1}{s!} \varphi_{\mu_1...\mu_s}\left(x\right) u^{\mu_1}...u^{\mu_s}, \label{Asymgen}
\end{equation}
where we have introduced the constant $(d+1)$-dimensional auxiliary vector $u^{\mu}$. The action of the covariant derivative also gets modified when acting on fields expressed as generating functions \eqref{Asymgen}, owing to the viel-bein dependence
\begin{align}
   \nabla_\mu \rightarrow \nabla_\mu+\omega_\mu^{ab}\, u_a\,\tfrac{\partial}{\partial u^b}, \quad 
  \left[\nabla_{\mu},\nabla_{\nu}\right] =
		\Lambda(u_\mu\partial_{u_\nu}-u_\nu\partial_{u_\mu})
		+R^{\Lambda}_{\mu\nu\rho\sigma}(x)u^\rho\partial_{u_\sigma}.
\end{align}
$\omega_\mu^{ab}$ is the spin-connection and viel-bein $e^a_{\mu}\left(x\right)$, with $u^a = e^a_{\mu}\left(x\right)u^{\mu}$. As a consequence of the vielbein postulate, we have
\begin{align}
	[\nabla_\mu,u^\nu] \;=\;0, \qquad
	[\partial_{u^\mu},\nabla_\nu] \; = \; 0.
\end{align}
$R^{\Lambda}_{\mu\nu\rho\sigma}$ is the Riemann tensor minus its 
constant trace part:
\begin{equation}
	R^{\Lambda}_{\mu\nu\rho\sigma} = 
		R_{\mu\nu\rho\sigma} 
		- \Lambda (g_{\mu\rho}g_{\nu\sigma} - g_{\nu\rho}g_{\mu\sigma}).
\end{equation}

\noindent In this framework, tensor operations are translated into an operator calculus, which simplifies manipulations significantly. The operations: box, divergence, symmetrised gradient, divergence, trace, symmetrised metric, and spin can be represented by the following operators:
\begin{align}
	\textrm{box:  } 			& (\nabla \cdot \partial_u)(\nabla \cdot u) , & 
	\textrm{divergence:  } 		& \nabla\cdot\partial_u,&
	\textrm{sym.~metric:  } 	& u^2, \nonumber \\
	\textrm{sym.~gradient:  } 	& u\cdot\nabla,	&
	\textrm{trace:  } 			& \partial_u^2,& 
	\textrm{spin:  } 			& u\cdot\partial_u .
\end{align}

\noindent As an exercise, reformulate the Fierz-system \eqref{fp} in the language of generating functions. The linearised gauge transformation \eqref{linhsg} takes the form
\begin{align} \label{gengi}
 \delta_{\xi}\varphi\left(x,u\right) = u \cdot \nabla \xi_{s-1}\left(x,u\right), \quad 
 \xi_{s-1}\left(x,u\right) = \frac{1}{(s-1)!}\xi_{\mu_1...\mu_{s-1}}u^{\mu_1}...u^{\mu_{s-1}}.
\end{align}
Using the commutators
\begin{subequations}
\begin{align}
	[\Box,u\cdot\nabla]=& \Lambda \left[u\cdot\nabla(2u\cdot\partial_u+d-1)-2u^2\nabla\cdot\partial_u\right]\\
					   &+2R^{\Lambda}_{\mu\nu\rho\sigma}\nabla^\mu u^\nu u^\rho\partial_{u_\sigma}-(\nabla_{\sigma} R^{\Lambda}_{\nu\rho} - \nabla_\rho R^{\Lambda}_{\nu\sigma})u^\nu u^\rho\partial_{u^\sigma}+R^\Lambda_{\nu\rho}u^\nu\nabla^\rho,\nonumber\\
	[\nabla\cdot\partial_u,u\cdot\nabla]&=\Box+\Lambda\left[u\cdot\partial_u(u\cdot\partial_u+d-2)-u^2\partial_u^2\right] +R^\Lambda_{\mu\nu\rho\sigma}u^\nu u^\rho \partial_{u_\mu}\partial_{u_\sigma}+R^\Lambda_{\mu\nu}u^\mu \partial_{u_\nu},
\end{align}
\end{subequations}
show that invariance of \eqref{divfp} under \eqref{gengi} implies \eqref{xifp}. Show that invariance of \eqref{casieomfp} fixes the mass of a spin-$s$ gauge field in AdS to be $\left(m_s R\right)^2 = (s-2)\left(s+d-2\right)-s$.\\

\noindent Hint: For AdS backgrounds $R^{\Lambda}_{\mu\nu\rho\sigma} = C_{\mu\nu\rho\sigma}=0$.

\end{framed}

\subsection{Lagrangian Formulation}
\label{subsec::lagform}
As a starting point in the quest for constructing interactions in a possible non-linear higher-spin action, we introduce the Lagrangian formulation of the free equations of motion \eqref{fpgauge} for a bosonic spin-$s$ gauge field on AdS$_{d+1}$. This description was obtained by Fronsdal in 1978 \cite{Fronsdal:1978rb} (and together with Fang for half integer spin \cite{Fang:1978wz}). We do not delve far into the free Fronsdal formulation,\footnote{See, for instance, the reviews \cite{Sorokin:2004ie,Rahman:2013sta,Gomez:2013sfb,Rahman:2015pzl}.} only briefly reviewing here the pertinent details. 

The first step is to take the Fierz system (\eqref{fpgauge} and \eqref{xifp}) off-shell, while keeping the correct number of physical degrees of freedom to describe a ${\cal D}\left(s+d-2,s\right)$ module. 
Towards deriving the complete on-shell system from a single equation, one deforms the Klein-Gordon equation \eqref{casi2} with divergence and trace terms 
\begin{equation}\label{eomdefo}
 \left[\Box- m^2_s + \alpha_1\left(u,\nabla\right)(\nabla\cdot\partial_u) + \alpha_2\left(u,\nabla\right) (\partial_u\cdot \partial_u)\right]\varphi_{s}\left(x,u\right) = 0,
\end{equation}
to account for the divergence \eqref{divg2} and trace \eqref{trg2} conditions. The guiding principle to determine the differential operators $\alpha_i\left(u,\nabla\right)$ is gauge invariance: Demanding that the deformations are at most two-derivative fixes
\begin{align}
  \alpha_1\left(u,\nabla\right) & = - \left(u \cdot \nabla \right), \\
  \alpha_2\left(u,\nabla\right) & = -u^2+\frac{1}{2}\left(u \cdot \nabla \right)^2,
\end{align}
but with the additional proviso that the gauge parameter is  \emph{traceless},
\begin{equation}\label{ginvfr}
 \delta_{\xi}\varphi_{s}\left(x,u\right) = u \cdot \nabla \xi_{s-1}\left(x,u\right), \qquad \left(\partial_u \cdot \partial_u\right)\xi_{s-1}\left(x,u\right) = 0.
\end{equation}
This also leads to a constraint on the field $\varphi_s$: Its double-trace is invariant under the gauge transformation \eqref{ginvfr} and so by unitarity must be set to zero, $\left(\partial_u \cdot \partial_u\right)^2 \varphi_{s}\left(x,u\right) = 0$.

One may verify that the Fronsdal formulation carries the correct number of physical degrees of freedom to describe a spin-$s$ gauge field, reducing to the Fierz system upon gauge fixing -- see e.g. \cite{Rahman:2015pzl} for details.

The algebraic trace constraints in the Fronsdal formulation
\begin{equation}\label{trcon}
 \left(\partial_u \cdot \partial_u\right)^2 \varphi_{s}\left(x,u\right) = 0, \qquad \left(\partial_u \cdot \partial_u\right)\xi_{s-1}\left(x,u\right) = 0,
\end{equation}
may seem unappealing, but they non-the-less achieve the goal of removing the derivative constraints, taking the Fierz system \eqref{fpgauge} off-shell. 
Forgoing the constraints \eqref{trcon} simply shifts the unconventional features elsewhere, such as: into additional auxiliary fields \cite{Pashnev:1997rm,Pashnev:1998ti,Burdik:2000kj} or introducing non-localities \cite{Francia:2002aa,Francia:2002pt}, to lift the trace constraint on the gauge parameter. For the double-trace condition on the field $\varphi_s$, to kill the non-unitary modes mentioned earlier one may try to impose appropriate boundary conditions. This is possible in flat space \cite{Francia:2010ap} though the situation in AdS, or whether this approach is compatible with introducing a source, is less clear. For these reasons, we stick to the Fronsdal formulation with algebraic constraints \eqref{trcon}.

The equation of motion \eqref{eomdefo} can be written in the form
\begin{equation}
 {\cal F}_{s}(x,u,\nabla,\partial_u)\varphi_{s}\left(x,u\right) = 0,
\end{equation}
where ${\cal F}_{s}$ is the so-called Fronsdal operator
\begin{align} \label{Fronsdaltensor}
{\cal F}_{s}(x,u,\nabla,\partial_u)
& =
\Box- m^2_s-u^2(\partial_u\cdot \partial_u)
-\;(u\cdot \nabla)\left((\nabla\cdot\partial_u)-\frac{1}{2}(u\cdot \nabla)
(\partial_u\cdot \partial_u)
\right).
\end{align}
This can be derived from the free action
\begin{align}  \label{fronsdal0}
    S^{(2)}_{\text{AdS}}\left[\varphi_s\right] & = \frac{s!}{2}  \int_{\text{AdS}_{d+1}}  \varphi_{s}\left(x;\partial_u\right) {\cal G}_s\left(x;u\right),
\end{align}
where ${\cal G}_s$ generalises to spin-$s$ gauge fields the linearised Einstein tensor
\begin{align}
   {\cal G}_s\left(x;u\right) & = \left(1-\frac{1}{4} \,u^2\, \partial_u \cdot \partial_u \right) \mathcal{F}_{s}\left(x; u, \nabla, \partial_u \right) \varphi_s\left(x, u\right).
\end{align}
Note that, crucially, the double-traceless condition on $\varphi_s$ ensures that the Bianchi identity is satisfied
\begin{equation}
   \left( \partial_u \cdot \nabla \right)\mathcal{G}_{s}\left(x, u\right) = 0.
\end{equation}

With the free action of a totally symmetric spin-$s$ gauge field on AdS$_{d+1}$, naturally the next step is to ask if we can construct interactions. Like for the determination of the kinetic term \eqref{fronsdal0}, this search is underpinned by the requirement of gauge invariance, and has been subject to decades of intense efforts. So far this approach has led to results for all possible cubic interactions \cite{Fradkin:1986qy,Fradkin:1987ks,Manvelyan:2004mb,Fotopoulos:2007yq,Boulanger:2008tg,Zinoviev:2008ck,Manvelyan:2009tf,Fotopoulos:2010nj,Bekaert:2010hk,Alkalaev:2010af,Boulanger:2011qt,Boulanger:2011se,Vasilev:2011xf,Joung:2011ww,Joung:2012rv,Joung:2012fv,Taronna:2012gb,Joung:2012hz,Boulanger:2012dx,Lopez:2012pr,Joung:2013doa} that may appear in a non-linear higher-spin action. 

In the following section we introduce a recent alternative approach to studying higher-spin interactions, which employs the AdS/CFT correspondence. As we shall see, holography seems to naturally imply the existence of interacting higher-spin theories on an AdS background, and has the potential to push further the successes of more conventional methods mentioned in the previous paragraph.

\section{The AdS/CFT Correspondence and Higher Spins}

In its most general form, the AdS / CFT correspondence \cite{Maldacena:1997re,Gubser:1998bc,Witten:1998qj} is  a conjectured duality which can be elegantly formulated as a simple equation:

\begin{equation}
    \text{AdS$_{d+1}$ QG}\; \overset{?}= \;\text{CFT}_{d}. \label{adscft0}
\end{equation}

\begin{figure}[h]
  \centering
  \includegraphics[scale=0.4]{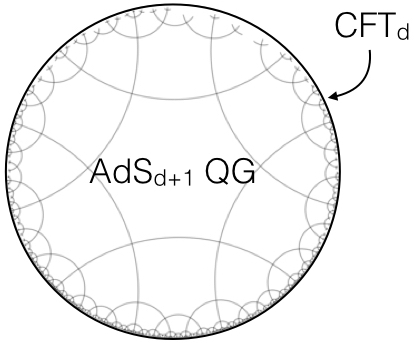}
  \caption{For a given dual pair \eqref{adscft0}, we can place the CFT$_{d}$ on the boundary (solid black boundary of disc) asymptotically AdS boundary of the dual gravity theory (entire disc). The grey curves are geodesics. This perspective can be obtained by taking Euclidean AdS, adding a point at infinity to the $\mathbb{R}^d$ boundary and compactifying it to $S^d$.} \label{hsduality}
\end{figure}

In words: Quantum Gravity\footnote{To be a bit more cautious we could say: Any theory that we
know how to define in the UV and behaves as ordinary gravity plus QFT in the infrared.} in asymptotically anti-de Sitter spacetime AdS$_{d+1}$ is postulated to be equal to a non-gravitational conformal field theory (CFT). This is known as a holographic duality, since the CFT lives in (at least) one lower dimension. Since the boundary of asymptotically AdS spaces are conformally flat, we can regard the CFT$_{d}$ as living on the `conformal boundary' of the dual theory in AdS$_{d+1}$. This is often depicted as in fig. \ref{hsduality}. 

In these lectures we are interested in a particular limit of the statement \eqref{adscft0}, in which higher-spin gauge fields are present in AdS$_{d+1}$. We thus won't delve into the details of this remarkable duality here,\footnote{Reviews of this vast topic include: \cite{Aharony:1999ti,Zaffaroni:2000vh,D'Hoker:2002aw,Polchinski:2010hw,Hubeny:2014bla}.}  covering only the salient concepts.

The equivalence \eqref{adscft0} is striking as, if true, it opens up the possibility to study gravity theories from the perspective of their CFT duals, and vice versa. Typically the dimensionless coupling $\lambda$ of the CFT$_{d}$ is related to the scale $\sqrt{\alpha^\prime}$ at which our gravity theory is sensitive to higher-derivative corrections via
\begin{equation}
    \lambda \;\sim\; \left(\frac{R^2}{\alpha^\prime}\right)^{d/2}. \label{param1}
\end{equation}
In a string theory context we have $\alpha^\prime = l^2_s$, the square of the string length. From the relationship \eqref{param1} we see that the holographic duality \eqref{adscft0} is strong-weak in nature, with two interesting limits:
\begin{enumerate}[label=\textbf{\arabic*}.]
    \item The point-particle limit: $\alpha^\prime/R^2 \rightarrow 0\,$, where the CFT coupling grows large $\lambda >> 1$.
    \item The high energy limit: $\alpha^\prime/R^2 \rightarrow \infty\,$, in which  $\lambda \rightarrow 0$.
\end{enumerate}
Although {\bf 1.} has been subject to intense study to mine the possibilities of investigating strongly coupled systems via relatively well understood General Relativity,\footnote{See \cite{Nastase:2015wjb,Ammon:2015wua} for pedagogical introductions.} {\bf 2.} underpins the topic of these lectures.

Why should the limit $\alpha^\prime/R^2 \rightarrow \infty\,$ be interesting? In this regime it is intuited that an infinite-dimensional symmetry may emerge, responsible for the good high energy behaviour of a UV-complete theory of gravity:\footnote{A well known example of this phenomenon is given by the Standard model of electro-weak interactions, where the massive $W^\pm$ and $Z$ bosons arise from the spontaneous breaking of an $SU\left(2\right) \times U\left(1\right)$ symmetry, which emerges at high energies.} To understand this expectation in more concrete terms, consider one of the most promising candidates for a complete theory of gravity: String Theory. Taking simply the open bosonic string in flat space, for the states on the first Regge trajectory, we have
\begin{equation}
    \alpha^\prime m^2_s = s-1, \qquad s=0, 1, 2, 3, ...\,.
\end{equation} 
In the $\alpha^\prime \rightarrow \infty\,$ limit we indeed recover a tower of higher-spin gauge fields in the spectrum, whose non-trivial interactions would generate an infinite-dimensional, higher-spin symmetry \cite{Gross:1988ue}. This phenomena can also be observed by considering higher-derivative counter-terms added to the Einstein-Hilbert action \cite{Camanho:2014apa}.

With a lot of symmetry comes a lot of control. In this regard, uncovering an infinite-dimensional higher-spin symmetry principle could shed some light on the elusive high-energy behaviour of gravity. This makes the limit {\bf 2.} even more profound, since, via holography, this highly symmetric phase of gravity can be probed through very simple, solvable, CFTs. In fact the emergence of higher-spin symmetry can also be seen from the dual CFT perspective: As we shall illustrate later, owing to the presence of a tower of conserved currents unbounded in spin in their spectrum, free CFTs are governed by an infinite-dimensional higher-spin symmetry. For the duality to hold, the theory in AdS should be governed by the same symmetry, making the existence of a highly symmetric phase of gravity even more plausible and more tractable to study.

In order to study this regime of holography in more detail, in the following we make the dictionary between the bulk and boundary theories more precise.

\subsection{The GKP/W Formula}
\label{subsec::gkpw}
In practice it is most convenient to formulate the holographic duality \eqref{adscft0} in terms of generating functions, in Euclidean signature. For concreteness, we work in Poincar\'e co-ordinates \eqref{poin}.

Let's start with the CFT side of the story. The generating function $F_{\text{CFT}}\left[{\bar \varphi}\right]$ of connected correlators in a CFT admits a path-integral representation, 
\begin{align}
    \exp\left(-F_{\text{CFT}}\left[{\bar \varphi}\right]\right) = \int D\phi\, \exp\left(-S_{\text{CFT}}\left[\phi\right] + \int d^dy\, {\bar \varphi}\left(y\right){\cal O}\left(y\right)\right). \label{qftgen}
\end{align}
 We use $\phi$ to collectively denote the fundamental field(s) in the theory, governed by the CFT action $S_{\text{CFT}}\left[\phi\right]$. The operator ${\cal O}$ is built from the fields $\phi$ in a gauge-invariant manner, and is sourced by ${\bar \varphi}\left(y\right)$. The source is not dynamical, rather a function that is fixed and under our control.

 Holography breathes life into the source ${\bar \varphi}$ of our CFT operators: Regarding the CFT$_d$ as living on the boundary of  AdS$_{d+1}$ (like in fig. \ref{hsduality}), the source is promoted to a fully fledged dynamical field $\varphi\left(y,z\right)$ in AdS governed by the bulk action $S_{\text{AdS}}\left[ \varphi \right]$. The only control we have over $\varphi$ is its boundary value ${\bar \varphi}\left(y\right)$ at $z=0$.

The GKP/W formula  \cite{Gubser:1998bc,Witten:1998qj,Klebanov:1999tb} puts the above holographic picture on a more concrete footing. It states that the physical quantity $F_{\text{CFT}}\left[{\bar \varphi}\right]$ in the CFT coincides with the AdS one $\Gamma_{\text{AdS}}\left[{\bar \varphi}\right]$, 
\begin{equation}
    F_{\text{CFT}}\left[{\bar \varphi}\right]=\Gamma_{\text{AdS}}\left[{\bar \varphi}\right],\label{gkpw}
\end{equation}
with 
\begin{align}
     \exp\left(-\Gamma_{\text{AdS}}\left[{\bar \varphi}\right]\right) & = \int_{\varphi|_{\partial \text{AdS}} =  {\bar \varphi}} {\cal D} \varphi \exp\left(-\frac{1}{G}S_{\text{AdS}}\left[\varphi\right]\right),
\end{align}
the bulk partition function and $G$ the gravitational constant. This identification and the precise boundary conditions are made more precise in the following section.

The GKP/W formula provides a prescription for computing correlation functions of gauge invariant operators in the CFT using the dual gravity theory. Connected correlation functions for instance can be obtained by functionally differentiating $\Gamma_{\text{AdS}}\left[{\bar \varphi}\right]$ instead:
\begin{equation}
    \langle {\cal O}_1\left(y_1\right) ... {\cal O}_n\left(y_n\right) \rangle_{\text{conn.}} = (-1)^n \frac{\delta}{\delta {\bar \varphi}_1\left(y_1\right)} ... \frac{\delta}{\delta {\bar \varphi}_n\left(y_n\right)} \Gamma_{\text{AdS}}\left[{\bar \varphi}_i\right]\Big|_{{\bar \varphi}_i=0}. \label{corgen0}
\end{equation}
In this holographic picture each functional derivative fires a $\varphi_i\left(y,z\right)$ particle into AdS.

At first sight the GKP/W formula doesn't appear to be so useful, since it involves dealing with the full gravity partition function. However, we may always take the limit in which gravity is classical. But what does this mean for the CFT? It turns out that quantum corrections in the bulk are sensitive to the number of degrees of freedom $N_{\text{dof.}}$ in the CFT, with\footnote{We can arrive to this by recalling that degrees of freedom in a CFT can be roughly measured by the overall coefficient ${\sf C}_{T}$ of the energy momentum tensor two-point function. From the GKP/W formula, we identify
\begin{equation}
\langle T T \rangle_{\text{CFT}} \quad \sim \quad {\sf C}_{T} \quad \sim \quad N_{\text{dof.}}  \qquad \leftrightarrow \qquad   \langle g g \rangle_{\text{AdS}} \quad \sim \quad \frac{R^{d-1}}{G} \quad \sim \quad \left(\frac{R}{\ell_p}\right)^{d-1},
\end{equation}
since, as we shall see explicitly in the following section, the CFT energy momentum tensor is dual to the graviton in AdS.} 
\begin{equation}
 \left(\frac{R}{\ell_p}\right)^{d-1} \quad \sim \quad N_{\text{dof.}}. \label{dofeq0}
\end{equation}
The classical limit in the bulk thus corresponds to having a large number of degrees of freedom in the dual CFT, $N_{\text{dof}}>>1$. In particular, the weak coupling $G << R$ expansion
\begin{equation}
 \Gamma_{\text{AdS}}\left[{\bar \varphi}\right] = \frac{1}{G}\Gamma_{\text{AdS}}\left[{\bar \varphi}\right]^{(0)} + \Gamma_{\text{AdS}}\left[{\bar \varphi}\right]^{(1)}+G \Gamma_{\text{AdS}}\left[{\bar \varphi}\right]^{(2)} + ...\,,\label{gammaexp}
\end{equation}
which is regulated by the dimensionless coupling
\begin{equation}\label{gndof}
   g := R^{1-d}G \: \sim \: 1/N_{\text{dof}}+{\cal O}\left(1/N^2_{\text{dof}}\right),
\end{equation}
translates to a large $N_{\text{dof}}$ expansion in the CFT, 
\begin{equation}
  F_{\text{CFT}} = N_{\text{dof}}\, F^{\left(0\right)}_{\text{CFT}} + F^{\left(1\right)}_{\text{CFT}} + \frac{1}{N_{\text{dof}}}\, F^{\left(2\right)}_{\text{CFT}} + ...\,.
\end{equation}
Given that the expansion \eqref{gammaexp} encodes the one-particle irreducible scattering amplitudes of the $\varphi_i$ in AdS, we encounter an elegant diagrammatic holographic interpretation of connected CFT correlators at large $N_{\text{dof}}$:
\begin{equation}\label{expfig}
  \includegraphics[scale=0.45]{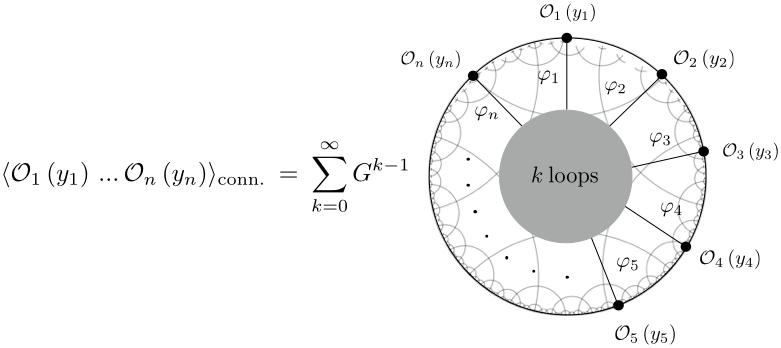}.
\end{equation}
The bulk diagrams on the right hand side are known as Witten diagrams. They were first computed in the context of the limit {\bf 1.} of the duality in \cite{Lee:1998bxa,Muck:1998rr,Mueck:1998iz,Freedman:1998bj,Chalmers:1998wu,Liu:1998ty,D'Hoker:1998mz,D'Hoker:1998gd,Liu:1998bu,Liu:1998th,D'Hoker:1999pj,D'Hoker:1999ea,Erdmenger:1999pz,Arutyunov:1999nw,Arutyunov:2000py,Bianchi:2003bd,Bianchi:2003ug,Uruchurtu:2007kq,Uruchurtu:2008kp,Penedones:2010ue,Uruchurtu:2011wh,Paulos:2011ie}. More recently Witten diagrams in the context of the limit {\bf 2.} where considered in 
\cite{Francia:2008hd,Giombi:2009wh,Giombi:2010vg,Chang:2011mz,Ammon:2011ua,Hijano:2013fja,Costa:2014kfa,Bekaert:2014cea,Bekaert:2015tva,Sleight:2016dba,Sleight:2016hyl}.
One-loop vacuum energies ($n=0$ and $k=1$ in \eqref{expfig}) have been computed in \cite{Giombi:2013fka,Giombi:2014iua,Giombi:2014yra,Beccaria:2014xda,Basile:2014wua,Pang:2016ofv,Bae:2016rgm,Bae:2016hfy,Gunaydin:2016amv,Bae:2017spv}.

\subsection*{The Field-Operator Map}

What does it take for a field $\varphi$ in an asymptotically AdS space to source an operator ${\cal O}$ of energy $\Delta$ and spin-$s$ in the dual CFT?
\begin{enumerate}
    \item The CFT operator must be gauge invariant, since the bulk physics is not sensitive to gauge dependent quantities $\rightarrow$ Spectrum of local operators are composed of traces.\footnote{In partciular, in the large $N_{\text{d.o.f.}}$ limit single-trace operators are identified with bulk single-particle states, while multi-trace operators are identified with multi-particle states in the bulk.} 
    \item The boundary value of $\varphi$ should have the same transformation properties under $SO\left(d,2\right)$ (quantum numbers) as the source to ${\cal O}$.
\end{enumerate}

How do we ensure condition 2? As we saw in \S \textcolor{blue}{\ref{subsec::uir}}, the quantum numbers of $SO\left(d,2\right)$ are energy and spin. By requiring invariance of \eqref{qftgen} under dilatations, the source ${\bar \varphi}$ of ${\cal O}$ has energy ${\tilde \Delta} = d-\Delta$.  We therefore have\footnote{See, for instance, \cite{Simmons-Duffin:2016gjk} for a derivation of the action of the dilation generator.}
\begin{equation}\label{bdye}
  \left[E,{\bar \varphi}\left(y\right)\right] = -i\left(d-\Delta+y \cdot \partial \right){\bar \varphi}\left(y\right),
\end{equation}
while on $\varphi$ it acts via a Lie derivative
\begin{equation}
  {\cal L}_{E}\,\varphi^{\mu_1...\mu_s}\left(z,y\right) = -i\left(z\partial_z+y \cdot \partial_y-s\right)\varphi^{\mu_1...\mu_s}\left(z,y\right).\label{bulke}
\end{equation}
By comparing \eqref{bulke} with \eqref{bdye}, to preserve invariance under conformal transformations as $z\rightarrow 0$ we need, for boundary directions $i$,
\begin{align}
    \varphi^{i_1 ... i_s}\quad \sim \quad z^{d+s-\Delta} \quad \text{as} \quad z \rightarrow 0, \label{sb}
\end{align}
where the coefficient $\varphi$ of $z^{d+s-\Delta}$ gives the source to ${\cal O}$.

A special case of the field-operator map occurs when the unitarity bound is saturated  (i.e. when $\Delta = s+d-2$), which is of interest from our perspective of higher-spin holography. Recall from \S \textcolor{blue}{\ref{subsec::uir}} that such representions of $SO\left(d,2\right)$ are known as short representations owing to the appearence of zero norm states in the Fock space, which should be factored out. For the bulk field $\varphi$, this multiplet shortening corresponds to the gauge invariance
\begin{equation}\label{gtunit}
  \delta_{\xi}\varphi_{\mu_1...\mu_s} = \nabla_{\left(\mu_1\right.}\xi_{\mu_2...\mu_s\left.\right).}
\end{equation}
In the CFT, this shortening manifests itself in the conservation of the lowest weight (primary) operator in the conformal multiplet:
\begin{equation}
  \partial^{i_1} {\cal O}_{i_1...i_s}\; \approx\; 0.
\end{equation}
Indeed, invariance of under the gauge variations \eqref{gtunit} corresponds to the conservation of ${\cal O}$:
\begin{equation}
 0 = \delta_{\xi} \int_{\partial \text{AdS}}d^dy\, {\bar \varphi}^{i_1...i_s}{\cal O}_{i_1...i_s} \overset{ \text{IBP}}= -\int_{\partial \text{AdS}}d^dy\, {\xi}^{i_2...i_s}\partial^{i_1}{\cal O}_{i_1...i_s}\; \implies \; \partial^{i_1}{\cal O}_{i_1...i_s} \approx 0.
\end{equation}
\\
\noindent In summary, the field-operator map tells us that:

\begin{equation}
    \includegraphics[scale=0.45]{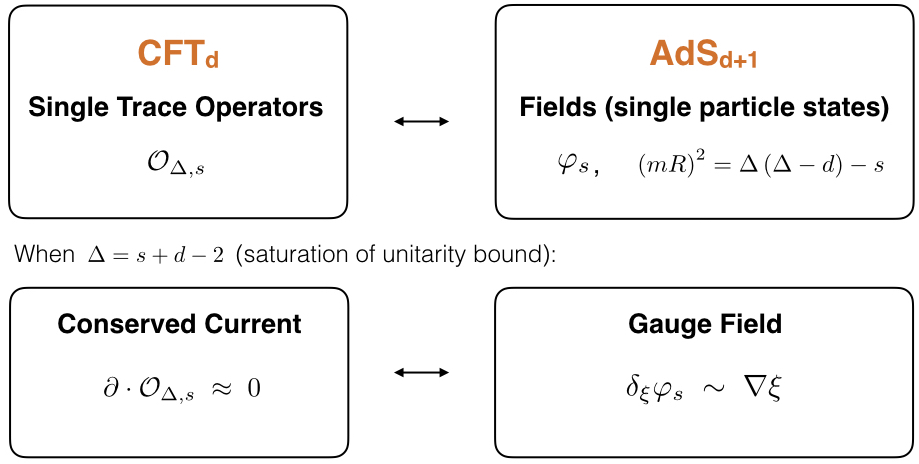}\nonumber
\end{equation}

\subsection{Higher Spin Holography}
\label{subsec::hsholo}
With the dictionary between CFT operators and bulk fields in place, let us explore in more detail holography in the limit {\bf 2}. Recall that this is the regime in which the dual CFT becomes \emph{free}.

For simplicity, consider a very basic example of a free CFT: A free massless $N$-component scalar in $d$-dimensions,
\begin{equation}
    S\left[\phi\right]=\frac{1}{2}\int d^dy \sum^{N}_{a=1} \partial_{i}\phi^a\partial^{i}\phi^a, \quad i=1,...,d,
    \end{equation}
    with equation of motion
 \begin{equation}
    \partial^2  \phi^a =0.    \label{eom}
        \end{equation}
This is known as the free scalar $O\left(N\right)$ vector model. For a free scalar we have $\Delta = \frac{d}{2}-1$, and throughout we work in Euclidean signature.

In the $O\left(N\right)$ singlet sector (as relevant for holography), amongst the single-trace operators we have the conserved and traceless\footnote{As required for a CFT.} stress-energy tensor
\begin{equation}
    T_{ij} = \partial_i \phi^a \partial_{j}\phi^a-\frac{1}{2}\delta_{ij} \partial^k\phi^a\partial_k\phi^a - \frac{d-2}{4\left(d-1\right)}\left(\partial_i\partial_j-\delta_{ij}\partial^2\right)\phi^a\phi^a, \label{tmu}
\end{equation}
where the right-most term is an improvement which is separately conserved, which is added to make the stress-energy tensor traceless. While it is straightforward to verify conservation using the equation of motion \eqref{eom}, notice that the stress energy tensor has dimension $\Delta=d$, and thus saturates the unitarity bound for a spin-2 lowest weight operator.

It is well known that the presence of a conserved and traceless stress-energy tensor is the signal of conformal invariance. But, free CFTs possess also much larger, higher-spin, symmetry. In this context, a higher-spin symmetry is typically\footnote{In certain dimensions higher-spin algebras have finitely many generators, such as $d=2$. However here we work in general $d$.} an infinite dimensional extension of the conformal algebra, generated by a tower of conserved charges that are unbounded in spin.\footnote{In fact in $d \ge 3$, assuming the existence of exactly one stress tensor, the presence of currents with spin $s \ge 3$ in a CFT implies that the theory is free \cite{Maldacena:2011jn,Boulanger:2013zza,Alba:2013yda,Alba:2015upa,Friedan:2015xea}.} 

That higher-spin charges are present in the free scalar $O\left(N\right)$ model can be seen explicitly by considering the generating function \cite{Bekaert:2009ud}
\begin{equation}
    {\cal J}\left(y;q\right) = \phi^a\left(y+iq\right)\phi^a\left(y-iq\right) = \sum^\infty_{s=0}\frac{1}{s!} {\cal J}_{i_1...i_s}\left(y\right)q^{i_1}...q^{i_s},\label{gfj}
\end{equation}
which describes a tower of operators ${\cal J}_{i_1...i_s}$ of ranks $s=0,2,4,...\,$. It is straightforward to verify that \eqref{gfj} is conserved, $\partial_y \cdot \partial_q {\cal J}\left(y;q\right) \approx 0$, which in turn implies\footnote{Since the scalar $\phi^a$ is real, odd spin operators do not appear in \eqref{gfj}. See exercise 3.1.}
\begin{equation}
    \partial^{i_1}{\cal J}_{i_1...i_s} \approx 0, \quad  s = 2, 4, 6, ...\label{ccs}
\end{equation}
The theory is thus, as a consequence of Noether's theorem, governed by the higher-spin symmetry generated by the charges associated to the conserved currents \eqref{ccs}. Let us note that for $s=0$ we have instead the scalar singlet operator
\begin{equation}
    {\cal O} = \phi^a\phi^a,
\end{equation}
of dimension $\Delta = 2 \times \left(\frac{d}{2}-1\right)=d-2$. Together with the tower of conserved currents \eqref{ccs}, this comprises the entire single-trace sector of the theory. 

A would-be dual theory in AdS should also be governed by (an appropriately gauged form) of the same higher-spin symmetry. According to the field-operator map, we expect the single-particle spectrum on AdS$_{d+1}$ to consist of a tower of gauge fields $\varphi_s$ for each even spin $s=2, 4, 6, 8, ...$ corresponding to each of the conserved currents \eqref{ccs} in the CFT, and a parity even scalar $\varphi_0$ dual to the scalar single trace operator ${\cal O} = \phi^a\phi^a$ \cite{Sezgin:2002rt,Klebanov:2002ja}.

While the discussion above was just representation theory, let us note that since (as we shall see explicitly) the correlation functions of operators in the singlet sector are non-trivial, owing to the GKP/W formula \S \textcolor{blue}{\ref{subsec::gkpw}}, we expect non-trivial interactions amongst the higher-spin gauge fields on AdS$_{d+1}$. In this regard, the AdS/CFT correspondence appears to make the potential existence of consistent interacting theories of higher-spin gauge fields quite natural.

\begin{framed}
\noindent \underline{Exercise 3.1:} Higher Spin Conserved Currents
\\

\noindent Extract the explicit form of the spin-$s$ conserved currents from the generating function \eqref{gfj}\footnotemark
\begin{equation}
    {\cal J}_{i_1...i_s} = i^s\sum^s_{k=0}\left(-1\right)^k\binom{s}{k}\partial_{\left(\right.i_1}...\partial_{i_k}\phi^a\partial_{i_{k+1}}...\partial_{\left.i_s\right)}\phi^a.\label{explccs}
\end{equation}
Hint:
\begin{equation}
    \phi^a\left(y+iq\right)\phi^a\left(y-iq\right) = e^{i q \cdot \partial_x}\phi^a\left(x\right)e^{-i q \cdot \partial_x}\phi^a\left(x\right).
\end{equation}
Notice that the currents \eqref{explccs} are not traceless, but, just like for the stress-energy tensor is CFT, one can add improvements to make them traceless.

One direct way \cite{Craigie:1983fb} to obtain the improved current is to consider the generating function ansatz 
\begin{equation}
   {\cal J}_s\left(y;z\right) =  f^{(s)}(z\cdot\partial_{y_1},z\cdot\partial_{y_2})\phi^a(y_1)\phi^a(y_2)\big|_{y_1,y_2 \rightarrow y}\,\label{jf},
\end{equation}
with null auxiliary vectors $z^2=0$ enforcing tracelessness. By demanding that ${\cal J}_s$ is annihilated by the conformal boost
\begin{equation}
    P^-{\cal J}_s=0,
\end{equation}
as per the definition of a primary (lowest weight) CFT operator, show that 
\begin{equation}
\left[(\tfrac{d}{2}-1+y\,\partial_y)\partial_y +(\tfrac{d}{2}-1+{\bar y}\,\partial_{\bar y})\partial_{\bar y}\right]f^{(s)}(y,{\bar y})=0.
\end{equation}
The solution is given in terms of a Gegenbauer polynomial
    \begin{equation}
    f^{(s)}\left(y,{\bar y}\right) = \left(y+{\bar y}\right)^s\,C^{\left(\frac{d-2}{2}\right)}_s\left(\frac{y-{\bar y}}{y+{\bar y}}\right)\,.\label{ccf}
\end{equation}

Notice also that \eqref{explccs} is zero for odd spins $s$. Show that by considering instead a complex scalar, 
\begin{equation}
    {\cal I}\left(y;q\right) = \phi^a\left(y+iq\right)\phi^{* a}\left(y-iq\right),
\end{equation}
as in the $U\left(N\right)$ theory, one obtains a tower of conserved currents for each integer spin.
\end{framed}
\footnotetext{The currents \eqref{explccs} are proportional to those derived by Berends, Burgers and van Dam in \cite{Berends:1985xx}. See also \cite{Anselmi:1999bb,Vasiliev:1999ba,Konstein:2000bi,Gelfond:2006be}.} 

It turns out the spectrum of the conjectured dual bulk theory is precisely that of the so-called minimal bosonic Vasiliev higher-spin theory in AdS$_{d+1}$ \cite{Vasiliev:2003ev}, which was in fact developed (in AdS$_4$) before the emergence of AdS/CFT \cite{Vasiliev:1990en}. The theory is formulated at the level of the equations of motion,\footnote{See \cite{Boulanger:2011dd,Boulanger:2012bj,Didenko:2015pjo} for proposals for an action, using the machinery of auxiliary fields.} by introducing an infinite-dimensional auxiliary internal space.\footnote{See \cite{Bekaert:2005vh,Iazeolla:2008bp,Didenko:2014dwa} for reviews of the Vasiliev system.} Comparably little is known about higher-spin interactions without the auxiliary fields and in the `metric-like' formalism,\footnote{It is called `metric-like' because the metric is generalised by the totally symmetric tensor $\varphi_{\mu_1...\mu_s}$.} which would extend the free Fronsdal action \eqref{fronsdal0} to the interacting level. On the other-hand the GKP/W formula \eqref{gkpw} seems to provide a possible implicit definition of a non-linear on-shell action. In these lectures we shall demonstrate how to use the GKP/W formula to extract metric-like higher-spin interactions on AdS$_{d+1}$, free from auxiliary fields.

Let us emphasise that above we have considered the simplest instance of higher-spin holography, where in the bulk one has just a tower of gauge fields of increasing spin.\footnote{This is in contrast to the stacks of Regge trajectories like in String Theory, which is mirrored in the dual CFT description with the basic fields being in the adjoint representation, as opposed to the fundamental representation considered here.} One may question whether holography holds in this more general scenario. On the other hand, Vasiliev's theory may be considered as a toy-model for studying the first Regge trajectory of string theory on AdS$_{d+1}$ in the tensionless limit (which, as we have seen, indeed consists of a tower of higher-spin gauge fields). In this scenario, one is free to study the consequences of higher-spin symmetry without the added complication of coupling higher-spin gauge fields to matter. 
Recent efforts have been dedicated, with the guidance of holography, to embed Vasiliev's higher-spin theory into tensionless string theory \cite{Bianchi:2003wx,Beisert:2003te,Beisert:2004di,Bianchi:2004ww,Bianchi:2004xi,Bianchi:2005yh,Bianchi:2005ze,Bianchi:2006gk,Chang:2012kt,Gaberdiel:2014cha,Gaberdiel:2015mra,Gaberdiel:2015uca,Gaberdiel:2015wpo}. The matter fields belonging to subleading Regge trajectories are organised into multiplets of the higher-spin symmetry generated by the first Regge trajectory.

The set-up considered here is the simplest amongst AdS higher-spin / vector model dualities. In particular, the spectrum of single-particle states includes only \emph{even} spin gauge fields. There exists an extension  of this theory to one whose spectrum contains a gauge field for \emph{each} integer spin. The conjectured dual CFT description is the free $U\left(N\right)$ invariant theory of a \emph{complex} scalar field $\phi^a$.\footnote{See exercise 3.1.}
 Further instances of higher-spin vector model dualities have been postulated, including: The so-called type-B theory in AdS$_4$ with a parity odd scalar in the bulk, Chan-Paton factors and supersymmetry \cite{Leigh:2003gk,Sezgin:2003pt,Giombi:2011kc,Chang:2012kt}. For comprehensive reviews on the higher-spin / vector model dualities, see: \cite{Bekaert:2012ux,Giombi:2012ms,Giombi:2016ejx}.

To move towards our goal of studying higher-spin interactions from a holographic perspective, in the following sections we introduce some useful tools for dealing with higher-spin fields in the metric-like formalism.

\section{The Ambient Space Formalism}
\label{sec::ambient}

The ambient space formalism is an instrumental tool in handling spinning fields and operators in the context of AdS/CFT. Dating back to Dirac \cite{Dirac:1936fq}, the basic idea is that fields in AdS$_{d+1}$, and respectively dual operators in the CFT$_d$, may be represented by fields in a $\left(d+2\right)$-dimensional flat ambient space. This makes manifest the mutual bulk and boundary $SO\left(d,2\right)$ symmetry. For metric-like calculations in anti-de Sitter space, an attractive feature of the ambient space formalism is that expressions intrinsic to the AdS
manifold (e.g. involving non-commuting covariant derivatives) can be expressed in terms of simpler-flat
space ones of the ambient space (e.g. commuting partial derivatives). Owing to these
simplifying features, the ambient formalism has enjoyed a wide variety of applications in AdS, HS and conformal field theory. For a flavour, see for example: \cite{Mack:1969rr,Fronsdal:1978vb,Biswas:2002nk,Weinberg:2010fx,Costa:2011mg,Taronna:2012gb}.

We review the basic features of the ambient approach here, and refer the reader to \cite{Grigoriev:2011gp,Taronna:2012gb,Bekaert:2012vt} for further details. 

\subsection{Bulk Fields}
\label{subsec::buf}
Our goal is to formulate totally symmetric spin-$s$ unitary representations of the AdS isometry in the language of ambient space. A key point to keep in mind is that, in representing a field on AdS by a tensor living in the \emph{higher-dimensional} ambient space, one has to make sure that the number degrees of freedom is kept constant. 

In the ambient formalism a given smooth rank-$r$ field  $\,t_{\mu_1...\mu_r}\left(x\right)$ intrinsic to the AdS$_{d+1}$ manifold is assigned a representative $T_{A_1...A_r}\left(X\right)$ of the same rank, which lives in the flat $(d+2)$-dimensional ambient space and in the same representation of $SO\left(d,2\right)$. Naturally, the pull back of $T$ onto the AdS must satisfy
\begin{align}
i^{*}\: : \: T_{A_1...A_r}\left(X\right) \: \longmapsto \: t_{\mu_1...\mu_r}\left(x\right) = \frac{\partial X^{A_1}\left(x\right)}{\partial x^{\mu_1}}\, ...\, 
\frac{\partial X^{A_r}\left(x\right)}{\partial x^{\mu_r}} T_{A_1...A_r}\left(X\left(x\right)\right). \label{pullback}
\end{align}
However, at this level the choice of representative $T$ is not unique. Indeed, the $SO\left(d,2\right)$ generators \eqref{sod2} are interior to AdS, $\left[J_{AB},\,X^2+R^2\right]=0$, and are not sensitive to the extension of $T$ away from the AdS manifold. More explicitly, there are two sources of ambiguity:
\begin{enumerate}
        \item Addition of tensors with components perpendicular to the AdS manifold, which sit in the kernel of the pullback \eqref{pullback}: 
    \begin{equation}
X^2 = -R^2 \: \implies \: \frac{\partial X}{\partial x^{\mu}} \cdot X \: \bigg|_{X^2 = -R^2}\,= 0. 
\end{equation}
\item Dependence on the radial direction $\rho = \sqrt{-X^2}$.\footnote{In particular, we are free to multiply by $\rho/R$ where $\rho$ is the radial co-ordinate $\rho = \sqrt{-X^2}$, or add terms proportional to $X^2+R^2$.}
\end{enumerate}
A prescription to fix the above ambiguities was provided by Fronsdal \cite{Fronsdal:1978vb} in the `70s. For ambiguities of type 1, we impose that $T$ is tangent to submanifolds of constant $\rho$:\footnote{This condition can be imposed by hand, applying the projection operator
\begin{equation}
\mathcal{P}^{B}_{A} = \delta^{B}_{A} - \frac{X_{A} X^{B}}{X^2}, \label{oproj}
\end{equation}
which acts on ambient tensors as
\begin{align}
\left(\mathcal{P} T\right)_{A_1...A_r} \; := \; \mathcal{P}^{B_1}_{A_1}...\mathcal{P}^{B_r}_{A_r}T_{B_1 ...B_r}, \qquad X^{A_i} \left(\mathcal{P} T\right)_{A_1...A_i...A_r}=0.
\end{align}}
\begin{equation}
    X^{A_i}T_{A_1...A_i...A_r} = 0, \qquad i = 1, ..., r.
\end{equation}
For the radial dependence, a simple (and thus convenient) condition is to impose that on-shell $T$ is a harmonic function in the ambient space
\begin{equation}
    \partial^2_X T_{A_1...A_r}=0.\label{harmt}
\end{equation}
In order for (see exercise 4.1) $T$ to carry the correct representation of $SO\left(d,2\right)$, the harmonic condition \eqref{harmt} implies that it is homogeneous
\begin{equation}\label{homot}
    \left(X \cdot \partial_X - \mu\right) T_{A_1...A_r}=0, \qquad \text{i.e.} \quad  T_{A_1...A_r}\left(\lambda X\right) = \lambda^{-\mu}  T_{A_1...A_r}\left(X\right),
\end{equation}
where we may choose either $\mu=\Delta$ or $\mu=d-\Delta$.

The above discussion also extends to the covariant derivative. The ambient representative of the covariant derivative $\nabla_{\mu}$ corresponding to the Levi-Civita connection on AdS$_{d+1}$ is simply given by 
\begin{equation}
    \nabla_{A} = {\cal P}^{B}_{A}\frac{\partial}{\partial X^B},\label{ambcovd}
\end{equation}
and acts via
\begin{equation}
\nabla = {\cal P} \circ \partial \circ {\cal P}.
\end{equation}
The first projection ensures that we are acting tangent to the AdS manifold. For example,
\begin{equation}
\nabla_B T_{A_1...A_r} = {\cal P}^{C}_{A}{\cal P}^{C_1}_{A_1}...{\cal P}^{C_r}_{A_r}\frac{\partial}{\partial X^C}\left({\cal P}T\right)_{C_1...C_r}.
\end{equation}

\subsection{Boundary Fields}

The above discussion also extends to fields living on the AdS boundary. In the same way, a rank-$s$ field $f_{i_1...i_s}(y)$ on the AdS$_{d+1}$ boundary is represented in ambient space by a function $F_{A_1...A_s}(P)$ living on a $P^+=\text{const.}$ slice of the projective null cone \S \textcolor{blue}{\ref{subsec::confbo}}.

If $f_{i_1...i_s}$ is a symmetric spin-$s$ boundary field of energy $\Delta$, its representative $F_{A_1...A_s}(P)$ is also symmetric and satisfies
\begin{align}\label{tlessf}
 \eta^{A_1A_2}F_{A_1...A_s}(P) & = 0 \\ \label{scalef}
 F_{A_1...A_s}(\lambda P) & = \lambda^{-\Delta}F_{A_1...A_s}(P), \quad \lambda(y) > 0. \end{align}
Like for the ambient description of bulk fields in the previous section, we impose that $F_{A_1...A_s}(P)$ is transverse to the light-cone
\begin{equation}
 P^{A_1}F_{A_1...A_s}(P) = 0,  \label{tbound}
\end{equation}
However in this case, since $P^2=0$, there is an extra redundancy
\begin{align}\label{extred}
 & \hspace*{3.5cm} F_{A_1...A_s}(P) \rightarrow F_{A_1...A_s}(P) + P_{\left(A_1\right.}\Lambda_{\left. A_2 ... A_s \right)},\\
 & P^{A_1}\Lambda_{ A_1 ... A_{s-1}} = 0, \quad \Lambda_{ A_1 ... A_{s-1}}(\lambda P) = \lambda^{-(\Delta+1)}\Lambda_{ A_1 ... A_{s-1}}(P), \quad \eta^{A_1A_2}\Lambda_{A_1 ... A_{s-1}} = 0,
\end{align}
which, together with \eqref{tbound}, removes the extra two degrees of freedom per index of $F_{A_1...A_s}$.

The scaling behaviour \eqref{scalef} extends the definition of $F_{A_1...A_s}$ away from the $P^+=\text{const.}$ slice, with the homogeneity degree fixed by the fact that $P \rightarrow \lambda(y) P$ re-scales the metric on $P^+=\text{const.}$ by an overall factor -- i.e. it implements a conformal transformation. See \S \textcolor{blue}{\ref{subsec::confbo}}. The tracelessness condition \eqref{tlessf} follows from the tracelessness of $f_{i_1...i_s}$: We have 
\begin{equation}
    f_{i_1 ...i_r}\left(y\right) = \frac{\partial P^{A_1}\left(y\right)}{\partial y^{i_1}} ... \frac{\partial P^{A_r}\left(y\right)}{\partial y^{i_r}} F_{A_1...A_r}\left(P\left(y\right)\right). \label{Apullback}
\end{equation}
In taking the trace of $f_{i_1 ...i_r}\left(y\right)$, on the RHS we implement the contraction
\begin{equation}\label{uplift}
    \delta^{ij}\frac{\partial P^A}{\partial y^i} \frac{\partial P^B}{\partial y^j} = \eta^{AB} + P^A Q^B + P^B Q^A, \quad \text{where} \quad Q^A = (1,\mathbf{0},-1).
\end{equation}
This gives vanishing trace of $f_{i_1 ...i_r}$ owing to the tracelessness \eqref{tlessf} and transversality \eqref{tbound} of $F_{A_1...A_r}$.

\subsection{Generating Functions}

Like for intrinsic tensors (exercise 1.2), it is useful employ an operator formalism and encode the ambient representatives of high-rank tensors in generating functions. 

\subsection*{Bulk fields}

For ambient representatives \S \textcolor{blue}{\ref{subsec::buf}} of totally symmetric bulk fields, we have
 \begin{equation}
 T_{A_1...A_s}\left(X\right) \: \longrightarrow \: T\left(X,U\right) = \frac{1}{s!} T_{A_1...A_s}\left(X\right) U^{A_1}...U^{A_s},
 \end{equation} 
with constant $(d+2)$-dimensional ambient auxiliary vector $U^{A}$. For \emph{traceless} fields, we may replace $U \rightarrow W$, with $W^2=0$.

Like for the intrinsic case, the covariant derivative \eqref{ambcovd} also gets modified in the generating function formalism. It takes the form
\begin{align}\label{ambcovdg}
\nabla_A = {\cal P}_{A}^{B} \frac{\partial}{\partial X^B} -\frac{X^B}{X^2} \Sigma_{AB}, \qquad X\cdot\nabla=0
\end{align}
where
\begin{equation}
\Sigma_{AB} = U_{\left[A\right.}\frac{\partial}{\partial U^{\left.B\right]}} = U_{A}\frac{\partial}{\partial U^{B}}-U_{B}\frac{\partial}{\partial U^{A}}, 
\end{equation}
is the spin operator in the ambient generating function formalism. 

In this case we have the operator algebra
\begin{align}\label{app::amsscomm}
    [X\cdot\partial_U,\nabla_A]&=0\,,& [\partial_U\cdot\partial_U,\nabla_A]&=0\,,&  [\nabla_A,X^2]&=0.
\end{align}

\begin{framed}
\noindent \underline{Exercise 4.1:} Homogeneity degree
\\

\noindent To demonstrate the power of the operator formalism, let's derive the homogeneity condition \eqref{homot} on harmonic \eqref{harmt} ambient representatives of totally symmetric fields.\\

\noindent Using that, in this totally symmetric case,
\begin{equation}
    iJ_{AB} = X_{\left[A\right.} \partial^X_{\left.B\right]} + U_{\left[A\right.} \partial^U_{\left.B\right]},
\end{equation}
derive the relation
\begin{equation}
    \frac{1}{2}J_{AB}J^{AB} = \left(U \cdot \partial_{U}\right)\left(U \cdot \partial_{U}+d-2\right)+ \left(X \cdot \partial_X\right)\left(d+X \cdot \partial_{X}\right)-X^2\partial^2_{X}.\label{casiuexp}
\end{equation}
For a field carrying the module ${\mathfrak so}(d,2)$ module ${\cal D}\left(\Delta,s\right)$ represented by the ambient tensor $T_{A_1...A_s}$, we have (see \S \textcolor{blue}{\ref{subsec::uir}})
\begin{equation}
   \frac{1}{2}J_{AB}J^{AB}T(X,U)  =  \left(\Delta\left(\Delta-d\right)+s\left(s+d-2\right)\right)T(X,U).
    \end{equation}
Using \eqref{casiuexp}, show that 
\begin{align}
  &  \partial^2_X  T(X,U) =0 \quad \implies \quad \left(X \cdot \partial_X -\mu\right)T(X,U) = 0, \quad \text{with} \quad \mu = \Delta \quad \text{or}\quad d-\Delta.
\end{align}
    
\end{framed}

\subsection*{Boundary fields}

For representatives \S \textcolor{blue}{\ref{subsec::confbo}} of traceless and totally symmetric boundary fields, we have 
\begin{equation}
F_{A_1 ...A_r}\left(P\right) \; \longrightarrow \; F\left(P,Z\right)=\frac{1}{r!}  F_{A_1 ...A_r}\left(P\right) Z^{A_1}...Z^{A_r}, \quad Z^2=0.
\end{equation}
The tangentiality condition \eqref{tbound}, expressed in the operator formalism as,
\begin{equation}
    \left(P \cdot \frac{\partial}{\partial Z}\right)F\left(P,Z\right) = 0,
\end{equation}
 can be enforced by demanding $F\left(P,Z+\alpha P\right) = F\left(P,Z\right)$, for any $\alpha$.  The extra redundancy \eqref{extred} is carried by the orthogonality condition $Z \cdot P=0$.
 
 The ambient auxiliary vector $Z$ is related to the intrinsic one $z$ (introduced in exercise 3.1) via
 \begin{equation}\label{Zz}
 Z^A = z^i \frac{\partial P^A}{\partial y^i} = z^i \left(y_i,\, \delta^{j}_i,\,- y_i\right) = \left(z \cdot y,\, z^j,\,-z \cdot y\right).
 \end{equation}

\section{Higher Spin Interactions from CFT}

Let's make more precise how we extract interactions in higher-spin gauge theories from CFT. Assuming the existence of a fully non-linear action principle $S_{\text{HS}}$ for a theory of higher-spin gauge fields, we perform a weak-field expansion around an empty AdS background in powers of the field fluctuations (which we denote collectively by $\varphi_i$)
\begin{equation}
 S_{\text{HS AdS}} = G S^{(2)}_{\text{HS AdS}}\left[\varphi_i\right] + G^{3/2}S^{(3)}_{\text{HS AdS}}\left[\varphi_i\right]+G^{2}S^{(4)}_{\text{HS AdS}}\left[\varphi_i\right]+...\,,\label{adsexp}
\end{equation}
where $S^{(n)}_{\text{HS AdS}}$ is order-$n$ in the field fluctuations about empty AdS. As we saw in \S \textcolor{blue}{\ref{subsec::lagform}}, the kinetic term of spin-$s$ gauge field $\varphi_s$ is given by the Fronsdal action 
\begin{align}  \label{fronsdal}
    S^{(2)}_{\text{HS AdS}}\left[\varphi_s\right] & = \frac{s!}{2}  \int_{\text{AdS}_{d+1}}  \varphi_{s}\left(x;\partial_u\right) \left(1-\frac{1}{4} \,u^2\, \partial_u \cdot \partial_u \right) \mathcal{F}_{s}\left(x; u, \nabla, \partial_u \right) \varphi_s\left(x, u\right),
\end{align}
where ${\cal F}_{s}$ is the Fronsdal operator 
\begin{align} \label{Fronsdaltensor2}
{\cal F}_{s}(x,u,\nabla,\partial_u)
& =
\Box- m^2_s-u^2(\partial_u\cdot \partial_u)
-\;(u\cdot \nabla)\left((\nabla\cdot\partial_u)-\frac{1}{2}(u\cdot \nabla)
(\partial_u\cdot \partial_u)
\right), \\ \nonumber
m^2_s R^2 & = \left(s+d-2\right)\left(s-2\right) - s.
\end{align}

The question is then the existence of non-trivial interaction terms $S^{(n)}_{\text{HS AdS}}$ with $n>2$. In the context of holography, from the GKP/W formula the possibility of consistent interacting theories of higher-spin gauge fields on an AdS background appears to be quite natural: 

We have \eqref{gkpw}
\begin{align}
  \exp\left(-F_{\text{free CFT}}\left[{\bar \varphi}\right]\right) =    \int_{\varphi|_{\partial \text{AdS}} =  {\bar \varphi}} {\cal D} \varphi \exp\left(-\frac{1}{G}S_{\text{HS AdS}}\left[\varphi\right]\right),
\end{align}
where $F_{\text{free CFT}}$ is the generating function of connected correlators in the dual free CFT.\footnote{In fact, for free CFTs the  $1/N_{\text{d.o.f}}$ expansion is \emph{exact},
\begin{equation}
  F_{\text{free CFT}} = N_{\text{dof}}\, F^{\left(0\right)}_{\text{free CFT}}.
\end{equation}}
From the relation \eqref{gndof} between the bulk coupling $g$ and $N_{\text{d.o.f}}$, at large $N_{\text{d.o.f}}$ we have
\begin{align} \label{adscft::ke}
    \exp\left(-F_{\text{free CFT}}\left[{\bar \varphi}\right]\right) = \exp\left(-\frac{1}{G}S_{\text{HS AdS}}\left[\varphi\right]\right)\Big|_{\varphi|_{\partial \text{AdS}} =  {\bar \varphi}} = \prod^\infty_{n=2} \exp\left(-\sqrt{G}^{\;n-2}S^{\left(n\right)}_{\text{HS AdS}}\left[\varphi\right]\right)\Big|_{\varphi|_{\partial \text{AdS}} =  {\bar \varphi}},
\end{align}
That $F_{\text{free CFT}}$ is non-trivial indicates, via \eqref{adscft::ke}, non-trivial interactions in the higher-spin gauge theory on AdS$_{d+1}$. With the knowledge of $ F_{\text{free CFT}}$, which is straightforward to determine in a free CFT, we may use \eqref{adscft::ke} to iteratively extract metric-like interactions in the dual higher-spin gauge theory:

\begin{equation}
\includegraphics[scale=0.65]{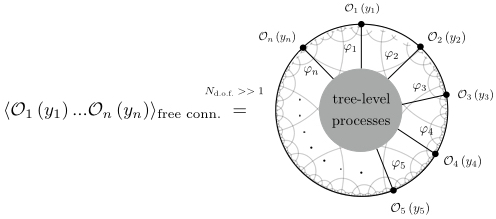}.\label{treelev}
\end{equation}

To this end a systematic approach needs to be developed to compute tree-level Feynman diagrams in AdS, known as Witten diagrams, for theories containing an infinite number of higher-spin gauge fields. This is the focus of the following sections.

\subsection{Witten Diagrams in Higher Spin Theories}

\subsubsection{Warm-up: Scalar Fields in AdS}
\label{subsubsfads}
For simplicity, for the remainder of these notes we set the AdS radius $R=1$.

To lay down the basics of evaluating tree-level Witten diagrams using the ambient space formalism, let's begin with the simplest example of a scalar field theory in AdS. In this way we are free from the extra complexity added when considering fields of higher-spin.

Consider the action

\begin{equation}\label{scalaract}
S\left[\varphi_i\right] = \frac{1}{G}\int_{\text{AdS}} \frac{1}{2} \nabla_{\mu}\varphi_i\nabla^{\mu}\varphi_i +m^2_i\varphi^2_i+ g \varphi_1\varphi_2\varphi_3+...\,, \quad i=1,2,3,
\end{equation}
with
\begin{equation}
m^2_i = \Delta_i\left(\Delta_i-d\right).
\end{equation}
In accordance with the GKP/W formula, at weak coupling the generating function of connected correlators at large $N_{\text{d.o.f}}$ in the dual CFT is given holographically by the on-shell action, subject to the boundary conditions
\begin{equation}
    \lim_{z \rightarrow 0} \varphi_i\left(z,y\right)z^{\Delta_i-d} \: = \: {\bar \varphi}_i\left(y\right) \label{bcscalar}.
\end{equation}
The first step is to solve the equations of motion,
\begin{equation}
    \frac{\delta S}{\delta \varphi_i} = \left(-\Box + m^2_i\right)\varphi_i + g \varphi_{j}\varphi_{k} + ... = 0, \qquad i \ne j \ne k \label{nlsc}
\end{equation}
subject to \eqref{bcscalar}. Since we are at weak coupling, this can be solved perturbatively in the boundary values ${\bar \varphi}$ using integral kernels. We expand
\begin{equation}
    \varphi_i\left(x\right) = \varphi^{(0)}_i\left(x\right) + \varphi^{(1)}_i\left(x\right)+ \varphi^{(2)}_i\left(x\right) + ...,
\end{equation}
where $\varphi^{(n)}_i$ is the solution at order $n+1$ in the ${\bar \varphi}$. This gives rise to the system of equations
\begin{align}
    \left(-\Box + m^2_i\right)\varphi^{(0)}_i & = 0, \\ \nonumber
    \left(-\Box + m^2_i\right)\varphi^{(1)}_i + g \varphi^{(0)}_j\varphi^{(0)}_k &= 0, \\\nonumber
    \left(-\Box + m^2_i\right)\varphi^{(2)}_i + g \varphi^{(0)}_j \varphi^{(1)}_k + g \varphi^{(1)}_j \varphi^{(0)}_k&= 0, \\ \nonumber
    & \hspace*{0.25cm} \vdots\,,
\end{align}
to be solved order-by-order in the ${\bar \varphi}$.

The solution of the first, linear, equation
\begin{equation}
    \left(-\Box + m^2\right) \varphi^{(0)}_i  = 0,
\end{equation}
can be constructed from the boundary data using the corresponding \emph{bulk-to-boundary propagator}. This the integral kernel
\begin{align}
    \varphi^{(0)}_i\left(z,y\right) & = \int_{\partial \text{AdS}} d^dy^{\prime}\, K_{\Delta_i}\left(z,y;{\bar y}\right){\bar\varphi}_i\left({\bar y}\right), 
\end{align}
where\footnote{The seemingly out of place factor of $2\Delta_i-d$ ensures consistency with the boundary limit of the bulk-to-bulk propagator \eqref{scbubu}.}
\begin{equation}
     \left(-\Box + m^2_i\right) K_{\Delta_i}\left(z,y;{\bar y}\right) = 0, \qquad \lim_{z \rightarrow 0}\left(z^{\Delta_i-d}K_{\Delta_i}\left(z,y;{\bar y}\right)\right) = \frac{1}{2\Delta_i-d} \delta^d\left(y-{\bar y}\right).\label{scalarbubo}
     \end{equation}
Higher-order solutions require the \emph{bulk-to-bulk propagator}
\begin{equation}
    \left(-\Box + m^2_i\right)\Pi_{\Delta_i}\left(x;x^\prime\right) = \frac{1}{\sqrt{|g|}}\delta^{d+1}\left(x-x^\prime\right). \label{scbubu}
\end{equation}
In this way we obtain the formal solution 
\allowdisplaybreaks
\begin{align} 
    \varphi^{(0)}_i\left(x\right) & = \int_{\partial \text{AdS}} d^dy^{\prime}\, K_{\Delta_i}\left(z,y;{\bar y}\right){\bar\varphi}_i\left({\bar y}\right), \\ \nonumber
   \varphi^{(1)}_i\left(x\right) & = - g \int_{\text{AdS}}d^{d+1}x^{\prime} \Pi_{\Delta_i}\left(x;x^\prime\right) \varphi^{(0)}_j\left(x^\prime\right)\varphi^{(0)}_k\left(x^\prime\right),\\ \nonumber
   \varphi^{(2)}_i\left(x\right) & =  - g \int_{\text{AdS}}d^{d+1}x^{\prime} \Pi_{\Delta_i}\left(x;x^\prime\right) \varphi^{(0)}_j\left(x^\prime\right)\varphi^{(1)}_k\left(x^\prime\right)- g \int_{\text{AdS}}d^{d+1}x^{\prime} \Pi_{\Delta_i}\left(x;x^\prime\right) \varphi^{(1)}_j\left(x^\prime\right)\varphi^{(0)}_k\left(x^\prime\right)\\ \nonumber
   & \hspace*{0.25cm} \vdots\,.
\end{align}
What remains is to insert the form of the kernels and perform the integration over AdS$_{d+1}$. The on-shell action is thus given by the diagrammatic expansion 
\begin{equation}
\includegraphics[scale=0.46]{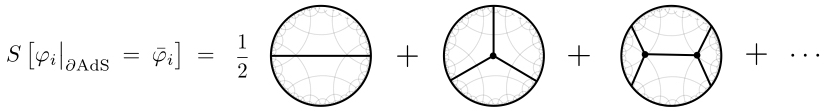}.\label{expsc}
\end{equation}
Connected correlation functions of the CFT operators ${\cal O}_i$ dual to the $\varphi_i$ can be computed at large $N_{\text{d.o.f}}$ by functionally differentiating \eqref{expsc} with respect to boundary values (sources) ${\bar \varphi}_i$.

In these lectures we restrict to the holographic computation of three-point functions at large $N_{\text{d.o.f}}$,
\begin{equation}
\includegraphics[scale=0.46]{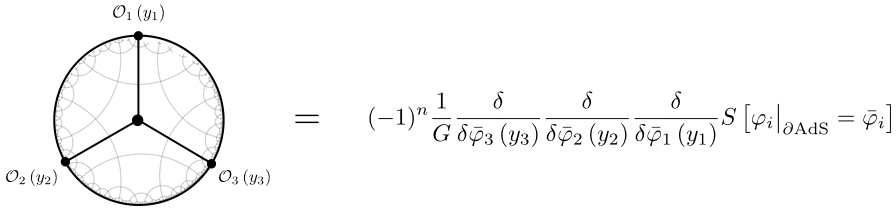}.\label{expsc2}
\end{equation}
which are generated by cubic interactions in AdS. To evaluate the corresponding Witten diagrams, in this case we just require the bulk-to-boundary propagators \eqref{scalarbubo}. In the Poincar\'e patch they are given by \cite{Witten:1998qj}
\begin{equation} \label{intprop}
    K_{\Delta}\left(z,y;{\bar y}\right) = C_{\Delta,0}\left(\frac{z }{z^2+\left(y-y^\prime\right)^2}\right)^{\Delta}, \qquad C_{\Delta,0} = \frac{\Gamma\left(\Delta\right)}{2\pi^{d/2}\Gamma\left(\Delta+1-\tfrac{d}{2}\right)},
\end{equation}
where the near-boundary behaviour \eqref{scalarbubo} fixes the overall coefficient.

\begin{framed}
\noindent \underline{Box 5.1:} Fixing the Propagator normalisation\\
 
\noindent Here we show explicitly how the propagator normalisation \eqref{intprop} is fixed by the near-boundary behaviour \eqref{scalarbubo}. At the linearised level, we have
\begin{equation}
 \varphi\left(z,y\right) = \int d^{d}{\bar y}\, K_{\Delta,0}\left(z,y;{\bar y}\right){\bar \varphi}\left({\bar y}\right).
\end{equation}

\noindent Setting for simplicity $y=0$ by translation invariance and going to radial co-ordinates $\rho = |{\bar y}|$, we have
\begin{align}
 \int d^{d}{\bar y}\, K_{\Delta,0}\left(z,0;{\bar y}\right){\bar \varphi}\left({\bar y}\right) & =  z^{\Delta} \Omega_{d-1} \int d\rho\, C_{\Delta,0}\frac{\rho^{d-1} }{\left(z^2+\rho^2\right)^{\Delta}} {\bar \varphi}\left(\rho\right) \\ \nonumber
 & = z^{d-\Delta} \Omega_{d-1} \int dt\,C_{\Delta,0}\frac{t^{d-1} }{\left(1+t^2\right)^{\Delta}} {\bar \varphi}\left(tz\right),
\end{align}
where $\Omega = \frac{2\pi^{d/2}}{\Gamma\left(d/2\right)}$ and in the second equality we made the change of variables $t = \rho/z$. Then, expanding ${\bar \varphi}\left(tz\right)$ about $z=0$,
\begin{align}
 \lim_{z \rightarrow 0} \varphi\left(z,0\right)z^{\Delta-d}  & =  C_{\Delta,0}\Omega_{d-1} \int dt\,\frac{t^{d-1} }{\left(1+t^2\right)^{\Delta}} {\bar \varphi}\left(0\right)\;+\; {\cal O}\left(z\right) \\ \nonumber
 & = C_{\Delta,0} \frac{\pi^{d/2}}{\Gamma\left(d/2\right)} \frac{\Gamma\left(d/2\right)\Gamma\left(\Delta-d/2\right)}{\Gamma\left(\Delta\right)}.
 \end{align}
 To obtain the boundary behaviour \eqref{bcscalar} we therefore require
 \begin{equation}
  C_{\Delta,0} = \frac{\Gamma\left(\Delta\right)}{2\pi^{d/2}\Gamma\left(\Delta+1-\tfrac{d}{2}\right)}.
 \end{equation}
 \end{framed}
In the following we demonstrate how to evaluate Witten diagrams \eqref{expsc2} using ambient space techniques \S \textcolor{blue}{\ref{sec::ambient}}, which admit a straightforward generalisation to the higher-spin case. In the ambient language, the propagator \eqref{intprop} takes the simple form 
\begin{equation}
    K_{\Delta}\left(X;P\right) = \frac{C_{\Delta,0}}{\left(-2X \cdot P\right)^{\Delta}},
\end{equation}
with bulk and boundary points 
\begin{equation}
    X = \left(\frac{z^2+y^2+1}{2z},\frac{y^i}{z},\frac{1-z^2-y^2}{2z}\right), \quad P=\left(\frac{1}{2}\left(1+{\bar y}^2\right),{\bar y}^i,\frac{1}{2}\left(1-{\bar y}^2\right)\right).
\end{equation}
At large $N_{\text{d.o.f.}}$, we have
\begin{equation}\label{k1k1k1}
    \langle {\cal O}_1\left(y_1\right){\cal O}_2\left(y_2\right){\cal O}_3\left(y_3\right) \rangle = g\int_{\text{AdS}} dX K_{\Delta_1}\left(X;P_1\right)K_{\Delta_2}\left(X;P_2\right)K_{\Delta_3}\left(X;P_3\right).
\end{equation}
Evaluating the bulk integral can be dramatically simplified by using the Schwinger-parameterised form for the propagator \cite{Penedones:2010ue,Paulos:2011ie}\footnote{\label{fn::gammaint} This is straightforward to obtain using the integral form of the Gamma function \begin{equation}
 \Gamma\left(t\right)\alpha^{-t} = \int^\infty_0 \frac{du}{u} u^t e^{-\alpha u}.
\end{equation}}
\begin{align}\label{schwing}
K_{\Delta}\left(X;P\right) = \frac{C_{\Delta,0}}{\Gamma\left(\Delta\right)} \int^\infty_0 \frac{dt}{t} t^\Delta e^{2t P \cdot X}.
\end{align}
In this way we have 
\begin{align}\label{3ptcont}
&g\int_{\text{AdS}} dX K_{\Delta_1}\left(X;P_1\right)K_{\Delta_2}\left(X;P_2\right)K_{\Delta_3}\left(X;P_3\right)\\ \nonumber
& \hspace*{3cm} = g\, \int^{\infty}_0 \prod\limits^{3}_{i=1}\left(\frac{C_{\Delta_i,0}}{\Gamma\left(\Delta_i\right)}\frac{dt_i}{t_i} t^{\Delta_i}\right) \int_{\text{AdS}} dX e^{2\left(t_1 P_1+t_2 P_2+t_3 P_3\right) \cdot X} \\ \nonumber
&  \hspace*{3cm} = g \pi^{\frac{d}{2}}\Gamma\left(\frac{- d + \sum\nolimits^3_{i=1} \Delta_i}{2}\right)  \int^{\infty}_0 \prod\limits^{3}_{i=1}\left(\frac{C_{\Delta_i,0}}{\Gamma\left(\Delta_i\right)}\frac{dt_i}{t_i} t^{\Delta_i}_i\right) e^{\left(-t_1t_2 P_{12} - t_1t_3 P_{13}-t_2t_3 P_{23}\right)},
\end{align}
where we used box 5.2 to evaluate the bulk integral and defined $P_{ij} = -2 P_i \cdot P_j$. Through the change of variables,
\begin{equation}
    t_1 = \sqrt{\frac{m_2m_3}{m_1}}, \quad t_2 = \sqrt{\frac{m_1m_3}{m_2}}, \quad t_3 = \sqrt{\frac{m_1m_2}{m_3}}, \label{tov}
\end{equation}
we obtain the final result 
\begin{align}\label{123scal}
& \langle {\cal O}_1\left(y_1\right){\cal O}_2\left(y_2\right){\cal O}_3\left(y_3\right) \rangle \\ \nonumber
& \hspace*{3cm} = \frac{1}{2} g\, \pi^{\frac{d}{2}}\Gamma\left(\frac{- d + \sum\nolimits^3_{i=1} \Delta_i}{2}\right) \int^{\infty}_0 \prod\limits^{3}_{i=1}\left(\frac{C_{\Delta_i,0}}{\Gamma\left(\Delta_i\right)}\frac{dm_i}{m_i} m_i^{\frac{\Delta_i}{2}}\right) \exp\left(-m_i P_{jk}\right) \\ \nonumber
& \hspace*{3cm} = g\, {\sf C}\left(\Delta_1,\Delta_2,\Delta_3; 0\right)  \frac{1}{P^{\frac{\Delta_1+\Delta_3-\Delta_2}{2}}_{13}P^{\frac{\Delta_2+\Delta_3-\Delta_1}{2}}_{23}P^{\frac{\Delta_1+\Delta_2-\Delta_3}{2}}_{12}}.
\end{align}
where we introduced 
\begin{align}
   & {\sf C}\left(\Delta_1,\Delta_2,\Delta_3; 0\right) \\ \nonumber
   & \hspace*{0.75cm} = \;\frac{1}{2}\pi^{\frac{d}{2}}\Gamma\left(\frac{- d + \sum\nolimits^3_{i=1} \Delta_i}{2}\right) C_{\Delta_1,0}C_{\Delta_2,0}C_{\Delta_3,0} \frac{\Gamma\left(\frac{\Delta_1+\Delta_2-\Delta_3}{2}\right)\Gamma\left(\frac{\Delta_1+\Delta_3-\Delta_2}{2}\right)\Gamma\left(\frac{\Delta_2+\Delta_3-\Delta_1}{2}\right)}{\Gamma\left(\Delta_1\right)\Gamma\left(\Delta_2\right)\Gamma\left(\Delta_3\right)}.
\end{align}

\begin{framed}
\noindent \underline{Box 5.2:} Tree-level contact diagrams using Schwinger Parameterisation \\

\noindent Consider the $n$-point contact diagram generated by the vertex involving scalars $\varphi_i$
\begin{equation}
 V_{12...n} = g \varphi_1 \varphi_2 ...\,\varphi_n,
\end{equation}
Using the Schwinger-parameterised form of the bulk-to-boundary propagators \eqref{schwing} one encounters the bulk integral
\begin{equation}\label{contn}
 {\cal A}^{\text{cont.}}\left(y_1,\,...\,,y_n\right) = \left(\prod^n_{i=1} C_{\Delta_i,0}\right) \int^{+\infty}_0 \prod^n_{i=1}\left(\frac{dt_i}{t_i} t^{\Delta_i}\right)\int_{\text{AdS}}dX \exp\left(2\sum\limits^n_{i=1}t_i\,P_i \cdot X \right),
\end{equation}
which is the extension of \eqref{3ptcont} to $n>3$. Defining $T = \sum^n_{i=1}t_iP_i$, by Lorentz invariance we may simply choose $T = |T|\left(1,\mathbf{0},0\right)$. Like this we obtain 
\begin{align}
\int_{\text{AdS}}dX \exp\left(2\sum\limits^n_{i=1}t_i\,P_i \cdot X \right) & =  \int^{+\infty}_0 \frac{dz}{z} z^{-d} \int d^{d}y\, e^{-\left(1+z^2+y^2\right)|T|/z} \\
& =  \pi^{d/2} \int^{+\infty}_0 \frac{dz}{z} z^{-d/2}\, e^{-\left(z-T^2/z\right)},
\end{align}
where in the second line we evaluated the Gaussian integral over $y$.\footnotemark \,Returning to the full \eqref{contn} and rescaling $t_i \rightarrow t_i/\sqrt{z}$, we can evalute the final integral over $z$ by using the integral representation of the Gamma function (see footnote \ref{fn::gammaint}.)
\begin{align}
&  \pi^{d/2} \int^{+\infty}_0 \prod^n_{i=1}\left(\frac{dt_i}{t_i} t^{\Delta_i}\right) \int^{+\infty}_0 \frac{dz}{z} z^{-d/2}\, e^{-\left(z-T^2/z\right)} \\ \nonumber
& \hspace*{4.5cm} = \pi^{d/2} \int^{+\infty}_0 \prod^n_{i=1}\left(\frac{dt_i}{t_i} t^{\Delta_i}\right) e^{T^2}\int^{+\infty}_0 \frac{dz}{z}\, z^{-\frac{d}{2}+\frac{1}{2}\sum\limits^n_{i=1}\Delta_i}\, e^{-z} \\ \nonumber
& \hspace*{4.5cm} = \pi^{d/2}  \Gamma\left(-\frac{d}{2}+\frac{1}{2}\sum\limits^n_{i=1}\Delta_i\right) \int^{+\infty}_0 \prod^n_{i=1}\left(\frac{dt_i}{t_i} t^{\Delta_i}\right)e^{-\sum\limits_{i < k} t_i t_k P_{ik}},\end{align}
where we used $T^2=\sum\limits_{i < k} t_i t_k P_{ik}$.

See, for instance, \cite{Penedones:2010ue,Paulos:2011ie,Fitzpatrick:2011ia} for other applications of this approach, in particular to establish Mellin representations of Witten diagrams.
\end{framed}
\footnotetext{In particular, \begin{equation}
 \int d^dy e^{-y^2|T|/z} = \left(\frac{z}{|T|}\right)^{d/2}\pi^{d/2}.
\end{equation}
}

\subsubsection{Witten Diagrams with External HS Fields}
\label{subsubwdehsf}
With the basics in place, we now turn to the case of external higher-spin fields. One of the main virtues of the ambient space approach is that it makes the $SO\left(d,2\right)$ symmetry manifest. That the ambient representative fields are homogeneous in both the bulk and boundary co-ordinates allows us to straightforwardly express a given Witten diagram with spinning external legs as some differential operator ${\cal F}\left(P_i,\partial_{P_i}\right)$ acting on a diagram with only external scalars, as illustrated below for tree-level three-point Witten diagrams
\begin{equation}\label{iterfig}
    \includegraphics[scale=0.44]{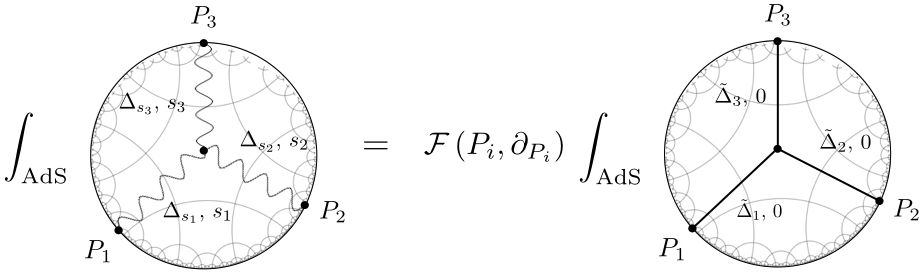}.
\end{equation}
In particular, we can reduce a three-point Witten diagram involving external spinning fields to one of the type \eqref{k1k1k1}, as we explain in the following.

\subsection*{Spinning Bulk-to-Boundary Propagators}

For a totally symmetric and traceless spin-$s$ field of energy $\Delta$, the corresponding bulk-to-boundary propagator satisfies the wave equation 
\begin{align}
    \left(-\Box + \Delta\left(\Delta-d\right)-s\right)
    K_{\Delta,\, \mu_1 ... \mu_s}{}^{i_1 ... i_s}\left(z, y;y^\prime\right) & = 0, \\
    \lim_{z \rightarrow 0}\left(z^{\Delta-d+s}K_{\Delta,\, \mu_1 ... \mu_s}{}^{i_1 ... i_s}\left(z, y;y^\prime\right)\right) & = \frac{\delta{}^{i_1 \, \ldots}_{\left\{\right.\mu_1 \, \ldots}\delta{}^{i_s }_{\mu_s\left.\right\}}}{2\Delta - d}\delta^{d}\left(y-y^\prime\right)\label{kdd}
\end{align}
The $SO\left(d,2\right)$ symmetry places constraints on the ambient representative
\begin{equation}
    K_{\Delta,s}\left(X,\alpha_1 W; \lambda P,\alpha_2 Z + \beta P\right) = \lambda^{-\Delta} \left(\alpha_1 \alpha_2 \right)^s K_{\Delta,s}\left(X, W; P, Z \right), \label{props}
\end{equation}
which fix its structure uniquely \cite{Mikhailov:2002bp}:\footnote{Since tracelessness in the bulk indices in this case is ensured by \begin{equation}
    P^2=0, \qquad Z^2=0, \qquad Z \cdot P = 0,
\end{equation}
we may replace $W \rightarrow U$, with $U^2 \ne 0$.} 
\begin{align}
    K_{\Delta,s}\left(X, W; P, Z \right) & = \left(W \cdot  {\cal P} \cdot Z \right)^s \frac{C_{\Delta,s}}{\left(-2 X \cdot P\right)^{\Delta}}.
    \end{align}
   The projector
    \begin{align}
     {\cal P}^A{}_B
     & = \delta^{A}_B - \frac{P^A X_B}{P \cdot X},\label{ps}
\end{align}
ensures transversality of the propagator at both its bulk and boundary points, while the overall coefficient $C_{\Delta,s}$ is fixed by equation \eqref{kdd}
\begin{equation}\label{poonorm}
  C_{\Delta,s} = \frac{\left(s+\Delta-1\right)\Gamma\left(\Delta\right)}{2 \pi^{d/2} \left(\Delta-1\right) \Gamma\left(\Delta+1-\tfrac{d}{2}\right)}.
\end{equation}
The latter also fixes the normalisation of the two-point function on the boundary
\begin{align}
\langle {\cal O}_{\Delta,s}\left(P_1\right){\cal O}_{\Delta,s}\left(P_2\right) \rangle\;&=\;K_{\Delta,s}\left(P_1;P_2\right)\\ \nonumber \\ \nonumber
&\includegraphics[scale=0.3]{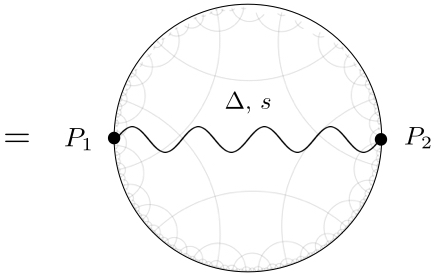}.
\end{align}

\subsection*{Spinning Witten diagrams from Scalar Witten diagrams}

The trick is that we can  express spinning propagators and their derivatives in the form
\begin{equation}
 \left( U_2 \cdot \partial_X \right)^n K_{\Delta,s}\left(X, U_1; P, Z \right) = {\cal G}\left(P,\partial_P;Z,U_1,U_2\right)K_{ \Delta+n,0}\left(X; P \right),\label{gprop}
\end{equation}
for some function ${\cal G}$ acting on a scalar bulk-to-boundary propagator, which we now determine. It is sufficient to consider the $n=0$ and $s=0$ cases:

For $n=0$ it is straightforward to confirm that
\begin{align} \label{bubos}
    & K_{\Delta,s}\left(X, U;P, Z \right) = \frac{1}{\left(\Delta-1\right)_s}\left({\cal D}_P\left(Z;U\right)\right)^s K_{\Delta,0}\left(X;P\right),
\end{align}
with 
\begin{equation}
 {\cal D}_P\left(Z;U\right) = \left(Z \cdot U\right)\left(Z \cdot \frac{\partial}{\partial Z} - P \cdot \frac{\partial}{\partial P}\right) + \left(P \cdot U \right)\left(Z \cdot \frac{\partial}{\partial P}\right),
\end{equation}
implementing the projection \eqref{ps}.

For $s=0$ we have 
\begin{equation}\label{partk}
\left( U \cdot \partial_{X} \right)^n K_{\Delta,0}\left(X; P \right)  = 2^n \left(\Delta\right)_n \left(U \cdot P\right)^n K_{\Delta+n,0}\left(X; P \right),
\end{equation}
which together with \eqref{bubos} fixes \eqref{gprop}. 

With the form \eqref{gprop} of bulk-to-boundary propagators and their derivatives, any Witten diagram with spinning external fields may be expressed in terms of one with only external scalars, with no derivatives acting upon them. For tree-level three-point Witten diagrams, this in particular means that we can recycle the result \eqref{123scal} we previously derived for the three-point Witten diagram generated by the basic cubic vertex \eqref{scalaract} for scalar fields. This is illustrated for a three-point Witten diagram involving a single external field of arbitrary mass and integer spin in the following.

\subsection*{Example: Three-point Witten Diagrams with Single HS Field}

We now demonstrate the approach described in the previous section for the simplest example of the tree-level three-point Witten diagram generated by a cubic interaction between two scalars $\phi_1$ \& $\phi_2$ of energies $\Delta_{1,2}$ and a totally symmetric spin-$s$ field $\varphi_s$ of energy $\Delta_s$ \cite{Costa:2014kfa,Bekaert:2014cea}.  This type of vertex is unique on-shell\footnote{I.e. unique when enforcing the constraints \eqref{fp} of the Fierz system, including the tracelessness and divergenceless constraints.} up to an overall coupling \cite{Metsaev:2005ar}, and takes the form\footnote{This can also be understood from the perspective of a dual CFT, since there is only a single conformally covariant tensor structure that can be written down for a correlation function containing two scalar operators and a spin-$s$ operator.}
\begin{equation}
    {\hat {\cal V}}_{0,0,s} = g \,\varphi_{\mu_1 ... \mu_s}\, \phi_1 \nabla^{\mu_1} ...  \nabla^{\mu_s} \phi_2 = s! g \,\phi_1(x) (\partial_u \cdot \nabla)^s \phi_2(x) \varphi_s\left(x,u\right), \label{basic}
\end{equation}
for some coupling $g$. In the second equality we expressed the vertex in the operator formalism (exercise 1.2).

\begin{figure}[h]
  \centering
  \includegraphics[scale=0.425]{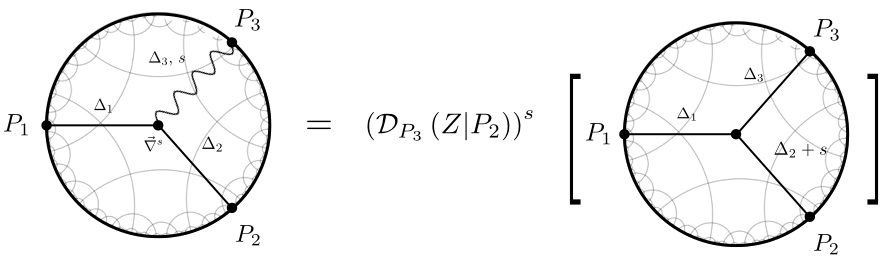}
  \caption{} 
 \label{3ptit}
\end{figure}

To apply the methods of the previous section to evaluate the corresponding tree-level Witten diagram, we have first to re-write the vertex \eqref{basic} in ambient space. Following the dictionary in \S \textcolor{blue}{\ref{sec::ambient}}, this is simply
\begin{equation}
    {\hat {\cal V}}_{0,0,s}(X) = g\,\varphi_{A_1 ... A_s}\, \phi_1 \nabla^{A_1} ...  \nabla^{A_s} \phi_2     
    = s! g\,\phi_1(X) (\partial_U \cdot \nabla)^s \phi_2(X) \varphi_s\left(X,U\right), \label{basicamb}
\end{equation}
where we recall the homogeneity and tangentiality conditions \S \textcolor{blue}{\ref{subsec::buf}} on ambient representatives. 

In fact in ambient space the vertex \eqref{basicamb} can be simplified further: Terms which deform the ambient representative \eqref{ambcovdg} of the covariant derivative from the $(d+2)$-dimensional flat partial derivative $\partial^A_X$ drop out, owing to the tracelessness and tangentiality of $\varphi_s(X,U)$,
\begin{equation}
    \partial_U \cdot \partial_U \varphi_s(X,U)=0, \qquad X \cdot \partial_U \varphi_s(X,U)=0.
\end{equation}
We therefore arrive to the very simple expression
\begin{equation}
    {\hat {\cal V}}_{s,0,0}(X) = s! g \phi_1(X) (\partial_U \cdot \partial_X)^s \phi_2(X) \varphi_s\left(X,U\right), \label{basicamb2}
\end{equation}
whose tree-level three-point Witten diagram we now evaluate following the approach outlined in the previous section. According to the standard Feyman rules \S \textcolor{blue}{\ref{subsubsfads}}, this is
\begin{align}
 &  \langle {\cal O}_{\Delta_1,0}(P_1){\cal O}_{\Delta_2,0}(P_2){\cal O}_{\Delta_3,s}(P_3;Z)\rangle\\ \nonumber
 &\hspace*{3.25cm} = s! g\int_{\text{AdS}}\, K_{\Delta_3,s}\left(X,\partial_U;P_3,Z\right)  K_{\Delta_1,0}\left(X;P_1\right) \left(U \cdot \partial_X\right)^s K_{\Delta_2,0}\left(X;P_2\right),
\end{align}
where the dual CFT operators ${\cal O}_{\Delta_1}$, ${\cal O}_{\Delta_2,0}$ and ${\cal O}_{\Delta_s,s}$ are sourced by $\phi_1$, $\phi_2$ and $\varphi_s$, respectively.

The core idea of \S \textcolor{blue}{\ref{subsubwdehsf}} was to generate such a Witten diagram with spinning external legs from the basic diagram \eqref{123scal} with only external scalars through the action of an appropriate differential operator in the boundary variables (equation \eqref{iterfig}). With the simple form of the vertex \eqref{basicamb2}, this is straightforward to attain using the expressions \eqref{bubos} and \eqref{partk} for the bulk-to-boundary propagators. We get,
\begin{align}
 &  \langle {\cal O}_{\Delta_1,0}(P_1){\cal O}_{\Delta_2,0}(P_2){\cal O}_{\Delta_3,s}(P_3;Z)\rangle\\ \nonumber
 &\hspace*{1.5cm} = \frac{s! g}{\left(\Delta_3-1\right)_s}\left({\cal D}_{P_3}\left(Z;\partial_U\right)\right)^s\int_{\text{AdS}}\, K_{\Delta_3,0}\left(X;P_3,Z\right)  K_{\Delta_1,0}\left(X;P_1\right) \left(U \cdot \partial_X\right)^s K_{\Delta_2,0}\left(X;P_2\right)
 \\ \nonumber
 &\hspace*{1.5cm} = \frac{s! \left(\Delta_2+1-\tfrac{d}{2}\right)_s}{\left(\Delta_3-1\right)_s}\left({\cal D}_{P_3}\left(Z;P_2\right)\right)^s g \int_{\text{AdS}}\, K_{\Delta_3,0}\left(X;P_3,Z\right)  K_{\Delta_1,0}\left(X;P_1\right) K_{\Delta_2+s,0}\left(X;P_2\right).
\end{align}
This relationship is depicted in figure \ref{3ptit}.

All that remains is to plug in the result \eqref{123scal} for the basic scalar three-point Witten diagram and evaluate the action of the differential operator ${\cal D}_{P}$, to arrive to the final expression\footnote{The action of ${\cal D}_P$ is given by
\begin{align}
&  \left({\cal D}_{P_3}\left(Z|P_2\right)\right)^s \;\frac{1}{P^{\frac{\Delta_1+\Delta_3-\Delta_2 - s }{2}}_{13}P^{\frac{\Delta_2+\Delta_3-\Delta_1+s}{2}}_{23}P^{\frac{\Delta_1+\Delta_2-\Delta_3+s}{2}}_{12}} \\ \nonumber 
 & \hspace*{4.5cm} = \left(\frac{\Delta_1 + \Delta_3-\Delta_2 - s}{2}\right)_s \frac{\left(\left(Z \cdot P_1\right)P_{23} - \left(Z \cdot P_2\right)P_{13}  \right)^s}{P^{\frac{\Delta_1+\Delta_3-\Delta_2 + s }{2}}_{13}P^{\frac{\Delta_2+\Delta_3-\Delta_1+s}{2}}_{23}P^{\frac{\Delta_1+\Delta_2-\Delta_3+s}{2}}_{12}}.
\end{align}} 
\begin{align}\label{oosfinito}
 & \langle {\cal O}_{\Delta_1,0}(P_1){\cal O}_{\Delta_2,0}(P_2){\cal O}_{\Delta_3,s}(P_3;Z)\rangle \\ \nonumber
 & \hspace*{4.5cm} =g{\sf C}\left(\Delta_1,\Delta_2,\Delta_3;s\right) \; \frac{\left(\left(Z \cdot P_1\right)P_{23} - \left(Z \cdot P_2\right)P_{13}  \right)^s}{P^{\frac{\Delta_1+\Delta_3-\Delta_2 + s }{2}}_{13}P^{\frac{\Delta_2+\Delta_3-\Delta_1+s}{2}}_{23}P^{\frac{\Delta_1+\Delta_2-\Delta_3+s}{2}}_{12}},
\end{align}
with \newpage
\begin{align}\label{00sW}
& {\sf C}\left(\Delta_1,\Delta_2,\Delta_3;s\right) \\ 
& \hspace*{0.25cm} = \frac{2^s \left(1-\frac{d}{2}+\Delta_2\right)_s}{\left(\Delta_3-1\right)_s} \left(\frac{\Delta_1 + \Delta_3-\Delta_2 - s}{2}\right)_s {\sf C}\left(\Delta_1,\Delta_2+s,\Delta_3;0\right) \nonumber \\ \nonumber
&  \hspace*{0.25cm} = C_{\Delta_1,0}C_{\Delta_2,0}C_{\Delta_3,s} \frac{2^s \pi^{\frac{d}{2}}\Gamma\left(\frac{\Delta_1+\Delta_2+\Delta_3-d+s}{2}\right)\Gamma\left(\frac{\Delta_1+\Delta_2-\Delta_3+s}{2}\right)\Gamma\left(\frac{\Delta_1+\Delta_3-\Delta_2+s}{2}\right)\Gamma\left(\frac{\Delta_2+\Delta_3-\Delta_1+s}{2}\right)}{2\Gamma\left(\Delta_1\right)\Gamma\left(\Delta_2\right)\Gamma\left(\Delta_3+s\right)}.
\end{align}
The result \eqref{oosfinito} has precisely the space-time dependence that is required by conformal symmetry (see \S \ref{subsubsec::csincft}). It can be expressed in more familiar intrinsic terms using that
\begin{align}
P_{ij} = \left(y_i-y_j\right)^2, \qquad Z^A = \left(z \cdot y_3,\, z^j,\,-z \cdot y_3\right), \qquad Z \cdot P_1 = z \cdot y_{13}, \qquad Z \cdot P_2 = z \cdot y_{23},  
\end{align}
to obtain 
\begin{align}\label{oosfinito2}
& \langle {\cal O}_{\Delta_1,0}(y_1){\cal O}_{\Delta_2,0}(y_2){\cal O}_{\Delta_3,s}(y_3;z)\rangle \nonumber \\
 & \hspace*{3cm} =g{\sf C}\left(\Delta_1,\Delta_2,\Delta_3;s\right) \; \frac{\left(\left(z \cdot y_{13}\right)y^2_{23} - \left(z \cdot y_{23}\right)y^2_{13}  \right)^s}{\left(y^2_{13}\right)^{\frac{\Delta_1+\Delta_3-\Delta_2 + s }{2}}\left(y^2_{23}\right)^{\frac{\Delta_2+\Delta_3-\Delta_1+s}{2}}\left(y^2_{12}\right)^{\frac{\Delta_1+\Delta_2-\Delta_3+s}{2}}}.
\end{align}

This approach is straightforward to extend to the general case with external fields with arbitrary mass and integer spin \cite{Sleight:2016dba}. The result \eqref{oosfinito2} itself is already enough to fix all higher-spin cubic interactions involving a single field of arbitrary integer spin, which we now consider.

\subsection{Holographic Reconstruction of HS Cubic Vertices}

Now that we know how to compute three-point Witten diagrams with an external field of arbitrary integer spin, we can already holographically re-construct part of the cubic order action of the minimal bosonic higher-spin theory  on AdS$_{d+1}$, which was introduced in \S \textcolor{blue}{\ref{subsec::hsholo}}. 

The GKP/W formula \S \textcolor{blue}{\ref{subsec::gkpw}} in this case reads:\footnote{In the present case of vector models, $N_{\text{d.o.f.}}=N$.}\footnote{Recall that for spin-$s$ gauge fields we have $m^2_sR^2= \left(s-2\right)\left(s+d-2\right)-s$.}
\begin{align} \label{fonc3pt}
   & F_{\text{free }O\left(N\right)}\left[{\bar \varphi}_s\right] \; \overset{N>>1}=\; \frac{1}{G}S_{\text{HS AdS}}\left[\varphi_s|_{\partial \text{AdS}}= {\bar \varphi_s}\right] \\ \nonumber
  & \hspace*{2cm}  = \int_{\text{AdS}_{d+1}} \sum_{s \in 2 \mathbb{Z}} \frac{s!}{2} \varphi_s\left(x,\partial_u\right)\left(\Box - m^2_s\right)\varphi_s\left(x,u\right) + \sum_{s_3\leq s_2\leq s_1} {\cal V}_{s_1,s_2,s_3}\left(\varphi_{s_i}\right) + ...\,
\end{align}
where $F_{\text{free }O\left(N\right)}$ is the generating function of connected correlators in the $d$-dimensional free scalar $O\left(N\right)$ model and $S_{\text{HS AdS}}$ is a would be non-linear action for the dual minimal bosonic higher-spin theory, expanded around AdS$_{d+1}$. Note that the action is on-shell, hence the dropping of gauge-dependent terms in the kinetic term from the off-shell Fronsdal action \eqref{fronsdal}.

In particular, the formula \eqref{fonc3pt} implies that cubic interactions between gauge fields of a given triplet $s_1$-$s_2$-$s_3$ of spins are fixed by the three-point function of conserved currents \eqref{jf} of the same triplet of spins,
\begin{align}\nonumber 
    \langle {\cal J}_{s_1}\left(y_1\right){\cal J}_{s_2}\left(y_2\right) {\cal J}_{s_3}\left(y_3\right)\rangle \;& \overset{N>>1}=\; \frac{\delta}{\delta {\bar \varphi_{s_3}}\left(y_3\right)} \frac{\delta}{\delta {\bar \varphi_{s_2}}\left(y_2\right)} \frac{\delta}{\delta {\bar \varphi_{s_1}}\left(y_1\right)}\int_{\text{AdS}_{d+1}} {\cal V}_{s_1,s_2,s_3}\left(\varphi_{s_i}\right)  \\ \nonumber \\ \label{solve123}
    & \includegraphics[scale=0.45]{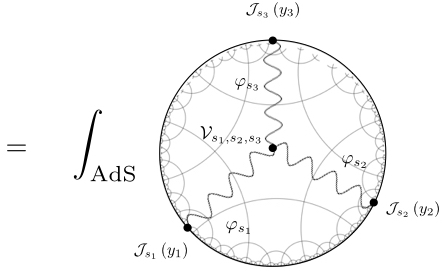}.
\end{align}
In practice one can proceed by making the most general ansatz for the vertex ${\cal V}_{s_1,s_2,s_3}$ and solving \eqref{solve123} for its form \cite{Sleight:2016dba}. This approach is particularly successful because the CFT is free, and so the correlation functions are straightforwardly computed by Wick contracting. 

For the ${\cal V}_{0,0,s}$ vertices, as discussed in \S \textcolor{blue}{\ref{subsubwdehsf}}, the ansatz one can write down is unique up to an overall coefficient. We can therefore use the result of the previous section to determine the cubic interactions involving a single gauge field of arbitrary spin, which we undertake in the following.

\subsection{Example: Cubic order action for $0$-$0$-$s$ interactions}
\label{subsubssec::00sex}
The simple illustrative example of this approach, which employs the results derived in \S \textcolor{blue}{\ref{subsubwdehsf}}, is to extract the
on-shell cubic interactions between two scalars $\varphi_0$ and a spin-$s$ gauge field $\varphi_s$ in the bulk action.

For consistency with the spin-$s$ gauge transformations, the ${\cal V}_{0,0,s}$ vertices have the form
\begin{equation}\label{00sv}
{\cal V}_{0,0,s} = g_{0,0,s} s! J_{s}\left(x,\partial_u\right) \varphi_s\left(x,u\right), \qquad \left(\partial_u \cdot \nabla\right) J_{s}\left(x,u\right)\; \approx\; 0,
\end{equation}
where $J_s$ is a spin-$s$ conserved current bi-linear in $\varphi_0$. Its explicit form is most straightforward to work out in ambient space \cite{Bekaert:2010hk}, which is demonstrated in exercise 5.1 below.

\begin{framed}
\noindent \underline{Exercise 5.1:}\; AdS Conserved Currents from Flat Space
\\

\noindent
Recall that we already encountered some spin-$s$ conserved currents that are scalar bi-linears in \S \textcolor{blue}{\ref{subsec::hsholo}}, but in flat space. In fact using the ambient space formalism, we can use this result to construct the analogous currents in AdS \cite{Bekaert:2010hk}:

Suppose that we have a symmetric rank-$s$ tensor $I_s$ in ambient space, that is conserved with respect to the ambient partial derivative
\begin{equation}\label{flambcon}
\left(\partial_U \cdot \partial_X\right) I_s(X,U) \; \approx\; 0.
\end{equation} 
Can $I_s$ also define a conserved current in AdS? Using ambient representative \eqref{ambcovdg} for the covariant derivative on AdS, show that 
\begin{align}\nonumber
 \left(\partial_U \cdot \nabla\right)  I_s\left(X,U\right) & = \left(\partial_U \cdot \partial_X\right)I_s\left(X,U\right) +\frac{U \cdot X}{X^2}\partial^2_UI_s\left(X,U\right) \\ 
& \hspace*{2cm} - \frac{X \cdot \partial_U}{X^2}\left[X \cdot \partial_X + U \cdot \partial_U +d\right]I_s\left(X,U\right)
\end{align}
The first term vanishes due to conservation \eqref{flambcon}, while the final term sits in the kernel of the pull-back \eqref{pullback} onto the AdS manifold and can thus be neglected. Therefore $I_s$ also represents a conserved current in AdS if 
\begin{equation}
\left[X \cdot \partial_X + U \cdot \partial_U +d\right]I_s\left(X,U\right) = 0.
\end{equation}
This condition is precisely satisfied by the flat space current \eqref{explccs} that we encountered in exercise 3.1, but with $\phi^a \rightarrow \varphi_0$:
\begin{equation}\label{ambcc}
I_s\left(X,U\right)=i^s\sum^s_{k=0}\left(-1\right)^k\binom{s}{k}\left(U \cdot \partial_X\right)^k\varphi_0\left(X\right)\left(U \cdot \partial_X\right)^{s-k}\varphi_0\left(X\right), 
\end{equation}
Where the ambient representative of the bulk scalar $\varphi_0$ satisfies (\eqref{harmt} and \eqref{homot})
\begin{align}
\partial^2_X \varphi_0\left(X\right)& = 0 \\
\left(X \cdot \partial_X - 2+d\right)\varphi_0\left(x\right)& = 0.
\end{align}
\end{framed}

Taking the expression \eqref{ambcc} for the conserved current, the vertex \eqref{00sv} in ambient space reads
\begin{align}\label{00sv2}
{\cal V}_{0,0,s} & = g_{0,0,s} s! i^s\,\varphi_s\left(X,\partial_U\right)\sum^s_{k=0}\left(-1\right)^k\binom{s}{k} \left(U \cdot \partial \right)^k \varphi_0\left(X\right) \left(U \cdot \partial \right)^{s-k} \varphi_0\left(X\right). 
 \end{align}
 On the other hand, recall in \S \textcolor{blue}{\ref{subsubwdehsf}} we argued that the structure of vertices involving two scalars and a spin-$s$ field is unique on-shell. In other words, it must be that
 \begin{equation}
 {\cal V}_{0,0,s} \; \approx\; \alpha {\hat {\cal V}}_{0,0,s},
 \end{equation}
 for some constant $\alpha$ and ${\hat {\cal V}}_{0,0,s}$ is the basic vertex \eqref{basic}. Indeed, integrating by parts and using the on-shell (Fierz-Pauli) conditions \eqref{fp} one finds
\begin{align}\label{00sv3}
{\cal V}_{0,0,s} & \approx\; 2^s {\hat {\cal V}}_{0,0,s} \;=\; g_{0,0,s} s! 2^s i^s\,\varphi_s\left(X,\partial_U\right) \left(U \cdot \partial \right)^{s} \varphi_0\left(X\right).
 \end{align}
 Re-cycling the result \eqref{oosfinito2} for the Witten diagram generated by the basic vertex ${\hat {\cal V}}_{0,0,s}$, from the bulk side we have \newpage
\begin{align}\label{bulk123res}
& \langle {\cal O}\left(y_1\right){\cal O}\left(y_2\right){\cal J}_s\left(y_3;z\right) \rangle \\ \nonumber
& \hspace*{2.5cm} \overset{N>>1}=  2^s g_{0,0,s}{\sf C}\left(d-2,d-2,s+d-2;s\right) \; \frac{\left(\left(z \cdot y_{13}\right)y^2_{23} - \left(z \cdot y_{23}\right)y^2_{13}  \right)^s}{\left(y^2_{13}\right)^{\frac{d}{2}-1+s}\left(y^2_{23}\right)^{\frac{d}{2}-1+s}\left(y^2_{12}\right)^{\frac{d}{2}-1}}.
\end{align}
What remains is to compare \eqref{bulk123res} with the result as computed in CFT, to which we now turn.

\subsubsection{Correlators in CFT}
\label{subsubsec::csincft}
In a CFT,\footnote{We only discuss the CFT side briefly here. For reviews / lecture notes on Conformal Field Theory see \cite{Qualls:2015qjb,Rychkov:2016iqz,Simmons-Duffin:2016gjk}.} the conformal symmetry fixes the structure of the three-point function up to a collection of coefficients \cite{Osborn:1993cr}. For the present case of three-point functions involving a single operator of non-zero spin, there is a unique structure compatible with conformal symmetry
\begin{equation}\label{cft3ptres}
\langle {\cal O}\left(y_1\right){\cal O}\left(y_2\right){\cal J}_s\left(y_3;z\right) \rangle \; = \; {\sf C}_{{\cal O}{\cal O}{\cal J}_s} \frac{\left(\left(z \cdot y_{13}\right)y^2_{23} - \left(z \cdot y_{23}\right)y^2_{13}  \right)^s}{\left(y^2_{13}\right)^{\frac{d}{2}-1+s}\left(y^2_{23}\right)^{\frac{d}{2}-1+s}\left(y^2_{12}\right)^{\frac{d}{2}-1}}.
\end{equation}
For free theories, ${\sf C}_{{\cal O}{\cal O}{\cal J}_s}$ is straightforward to compute by Wick contracting. To this end, it is convenient to employ the generating function representation \eqref{jf} for the spin-$s$ currents. For the free scalar $O\left(N\right)$ model one finds \cite{Diaz:2006nm} (see exercise 5.2)
\begin{align} \label{coo}
    {\sf C}_{{\cal J}_s {\cal O}{\cal O}} &  = 8 N \left[\frac{1+(-1)^s}{2}\right] \frac{2^s \left(\frac{d}{2}-1\right)_s  (d-3)_s}{\Gamma (s+1)}.
\end{align}

\begin{framed}
\noindent \underline{Exercise 5.2} Three-point function coefficient
\\

\noindent Using the generating function representation \eqref{jf} for the spin-$s$ currents, show that
\begin{align}\label{joo}
 &   \langle {\cal O}\left(y_1\right){\cal O}\left(y_2\right){\cal J}_s\left(y_3;z\right)\rangle \\ \nonumber
    & \hspace*{3cm} =  \frac{8N  }{\Gamma\left(\frac{d}{2}-1\right)^3}\left[\frac{1+(-1)^s}{2}\right]\left(z \cdot \partial_{y}+z \cdot \partial_{\bar y}\right)^sC^{\left(\frac{d-3}{2}\right)}_s\left(\frac{z \cdot \partial_{y}-z \cdot \partial_{\bar y}}{z \cdot \partial_{y}+z \cdot \partial_{\bar y}}\right) \\ \nonumber
    & \hspace*{3cm} \times \int^{\infty}_0 \left(\prod\limits^3_{i=1} \frac{dt_i}{t_i}t^{\frac{d}{2}-1}_i\right)\, e^{-t_1\,\left(y-y_1\right)^2-t_2\,\left(\bar y-y_2\right)^2-t_3\,\left(y_1-y_2\right)^2}\big|_{y,\bar y \rightarrow y_3}.
\end{align}
Hint: Express the two-point function of the fundamental scalar in Schwinger-parameterised form
\begin{equation}
    \langle \phi^a\left(y_1\right)\phi^b\left(y_2\right) \rangle = \frac{\delta^{ab}}{\Gamma\left(\frac{d}{2}-1\right)}\int^{\infty}_0 \frac{dt}{t}t^{\frac{d}{2}-1} e^{-t\, y^2_{12}}.\label{schwing2pt}
\end{equation}
By extracting the coefficient of $\left(y_{13} \cdot z \right)^s$ in \eqref{joo}, confirm the expression \eqref{coo} for the overall coefficient of the three-point function.
\end{framed}

Likewise, two-point functions are also fixed up to an overall coefficient by conformal symmetry
\begin{equation}
    \langle {\cal J}_{s}\left(y_1;z_1\right) {\cal J}_{s}\left(y_2;z_2\right)  \rangle = \frac{{\sf C}_{{\cal J}_s}}{\left(y^2_{12}\right)^{s+d-2}} \left(z_1\cdot z_2+\frac{2 z_1\cdot y_{12}\,z_2\cdot y_{21}}{y_{12}^2}\right)^s\,. \label{js2pt}
\end{equation}
A similar exercise in Wick contractions  yields \cite{Anselmi:1998bh,Diaz:2006nm},
\begin{equation}
 {\sf C}_{{\cal J}_s} =  2^{s+1} N \left[\tfrac{1+\left(-1\right)^s}{2}\right]\frac{(d-3)_s (d-3)_{2s}}{\Gamma\left(s+1\right)}.
\end{equation}

\subsubsection{Holographic reconstruction}
 Before we proceed to extract the cubic couplings $g_{0,0,s}$ holographically, let us emphasise that the GKP/W formula \eqref{gkpw} is only meaningful if the two-point function normalisations in both the bulk and boundary are consistent.  In the following we employ the unit normalisation\footnote{On the bulk side this entails sending ${\cal J}_{s} \rightarrow \frac{1}{\sqrt{C_{s+d-2,s}}}{\cal J}_{s}$, while in the CFT we send ${\cal J}_{s} \rightarrow \frac{1}{\sqrt{ {\sf C}_{{\cal J}_s}}}{\cal J}_{s}$.}
 \begin{equation}\label{unitn2pt}
    \langle {\cal J}_{s}\left(y_1;z_1\right) {\cal J}_{s}\left(y_2;z_2\right)  \rangle = \frac{1}{\left(y^2_{12}\right)^{s+d-2}} \left(z_1\cdot z_2+\frac{2 z_1\cdot y_{12}\,z_2\cdot y_{21}}{y_{12}^2}\right)^s.
 \end{equation} 
 To extract the cubic coupling we compare the bulk \eqref{bulk123res} and boundary \eqref{cft3ptres} results with normalisation \eqref{unitn2pt}, to obtain \cite{Bekaert:2015tva}\footnote{See also \cite{Petkou:2003zz} for the earlier $s=0$ case on AdS$_4$. The vertex in this case is in fact \emph{vanishing}, which can be seen by inserting $d=3$ and $s=0$ in \eqref{couplgend}. To reconcile this result with the \emph{non-zero} dual CFT correlator $\langle {\cal O} {\cal O} {\cal O} \rangle$, in this case one may add a boundary term to the bulk action which generates the CFT result. This was carried out \cite{Freedman:2016yue} for the duality with four-dimensional ${\cal N}=8$ gauged supergravity \cite{deWit:1982bul} in the bulk, which has no $A^3$ cubic couplings but the dual CFT correlators are non-vanishing.}
 
 \begin{align} \label{couplgend}
g_{0,0,s} =  \frac{2^{\tfrac{3d-s-1}{2}}\pi^{\tfrac{d-3}{4}}\Gamma\left(\frac{d-1}{2}\right)\sqrt{\Gamma\left(s+\tfrac{d}{2}-\tfrac{1}{2}\right)}}{\sqrt{N}\sqrt{s!}\,\Gamma\left(d+s-3\right)}. 
\end{align}
Holography therefore indicates that the sector of a would-be cubic order action on AdS$_{d+1}$ involving interactions of the gauge fields with two scalars (on-shell) takes the form
\begin{align}\label{c00sact}
&\frac{1}{G}S_{\text{HS AdS}}\left[\varphi_s\right] \\ \nonumber
& \hspace*{0.75cm} =  \sum_{s \in 2 \mathbb{Z}} \, \int_{\text{AdS}_{d+1}} \frac{1}{2} \varphi^{\mu_1...\mu_s}\left(x\right)\left(\Box - m^2_s\right)\varphi_{\mu_1...\mu_s}\left(x\right) + g_{0,0,s}\,2^s \varphi_{\mu_1...\mu_s} \varphi_0\left(x\right)\nabla^{\mu_1}...\nabla^{\mu_s}\varphi_0\left(x\right)+ ...\,,
\end{align}
with $m^2_s = \left(s-2\right)\left(s+d-2\right)$.

A few concluding comments:

\begin{itemize}

    \item In obtaining the result \eqref{c00sact}, we did not employ any notion of higher-spin \emph{gauge} symmetry -- under which the theory should still be invariant at the interacting level. An important non-trivial check of the result (together with its completion \cite{Sleight:2016dba} for the complete action at cubic order \eqref{fonc3pt} -- i.e. for any triplet of integer spins $s_1$-$s_2$-$s_3$) was the demonstration that it would coincide with the vertex obtained purely from requiring higher-spin gauge invariance at the interacting level \cite{Sleight:2016xqq}, i.e. the Noether procedure. This also served as a test of higher-spin holography itself, generalising the existing tree-level three-point function tests in AdS$_4$ \cite{Giombi:2009wh} and AdS$_3$ \cite{Chang:2011mz,Ammon:2011ua} to generic dimensions.
    
    \item At the same time, the holographic approach to constructing higher-spin interactions appears to be more efficient than the Noether procedure. Indeed, invariance under linearised higher-spin gauge transformations \eqref{linhsg} is insufficient to fix the relative couplings \eqref{couplgend}, requiring the consideration of higher-order consistency conditions. See e.g. \cite{Fradkin:1986qy,Vasilev:2011xf,Boulanger:2013zza,Kessel:2015kna}. The holographic approach requires only knowledge of three-point free CFT correlators.
    
    \item The methods and results presented here can be straightforwardly carried over to other instances of higher-spin holography. So far, the $0$-$0$-$s$ cubic couplings have also been holographically reconstructed for the higher-spin theory dual to the free fermion vector model \cite{Skvortsov:2015pea}. Moreover, 
    the tools and results \cite{Sleight:2016dba} presented here for evaluating three-point Witten diagrams    
    with external fields of arbitrary integer spin also apply to \emph{massive} fields and can be straightforwardly extended to representations of mixed-symmetry, as relevant for string theory. In particular, these methods and results are applicable beyond higher-spin-symmetric set-ups.
    
    \item The holographic reconstruction of interactions can also in principle be executed at quartic and higher-orders. So far, there have been results for the quartic self-interaction of the scalar in the minimal bosonic higher-spin theory \cite{Bekaert:2014cea,Bekaert:2015tva,Bekaert:2016ezc}. At this order the question of locality becomes important, as one is inevitably led to consider interactions with an unbounded number of derivatives. For investigations in this direction, see \cite{Barnich:1993vg,Taronna:2011kt,Dempster:2012vw,Vasiliev:2015wma,Boulanger:2015ova,Bekaert:2015tva,Skvortsov:2015lja,Taronna:2016ats,Bekaert:2016ezc,Vasiliev:2016xui,Taronna:2016xrm,Taronna:2017wbx,Roiban:2017iqg}.    
\end{itemize}

\bibliography{refs}
\bibliographystyle{JHEP}

\end{document}